\documentclass[twocolumn,twoside]{IEEEtran}

\ifCLASSINFOpdf

\else

\fi

\ifCLASSOPTIONcompsoc
    \usepackage[caption=false, font=normalsize, labelfont=sf, textfont=sf]{subfig}
\else
    \usepackage[caption=false, font=normalsize]{subfig}
\fi
\usepackage{lipsum}%
\usepackage[dvipsnames]{xcolor}
\usepackage{algorithm,algorithmic}
\usepackage{balance}
\usepackage{multicol}   
\usepackage{cite}
\usepackage{gensymb}
\usepackage{multirow}
\usepackage{graphics}
\usepackage{epsfig}
\usepackage{graphicx}
\usepackage{epstopdf}
\usepackage{textcomp}
\usepackage{amsmath}
\usepackage{mathtools}
\usepackage{filecontents}
\usepackage{lipsum,color}
\usepackage{amssymb}
\usepackage{float}
\usepackage{colortbl} 
\usepackage{times} 
\usepackage{amsthm}  

\usepackage{amsfonts}

\usepackage{arydshln}

\theoremstyle{break}

\begin{document}
\title{3D Uplink Channel Modeling of UAV-based mmWave Fronthaul Links for \\Future Small Cell Networks}


\author{ M.~T.~Dabiri,~and~M.~O.~Hasna,~{\it Senior Member,~IEEE}
	\thanks{Mohammad Taghi Dabiri and Mazen Omar Hasna are with the Department of Electrical Engineering, Qatar University, Doha, Qatar  (E-mail: m.dabiri@qu.edu.qa and hasna@qu.edu.qa).}
	}
\maketitle
\vspace{-1cm}
\begin{abstract}
In this study, we consider an unmanned aerial vehicle (UAV)-assisted heterogeneous network that is offered as a cost effective and easy to deploy solution to solve the problem related to transferring traffic of  distributed small cells to the core network. For any given distribution of small cell base stations (SBSs), we first characterize an accurate millimeter wave (mmWave) channel model for SBS to networked flying platform (NFP) by taking into consideration real parameters such as UAV's vibrations, distribution of SBSs, position of UAVs in the sky, real three-dimensional (3D) antenna pattern model provided by 3GPP along with interference caused by antenna side lobes and frequency reuse. Then, for the characterized channel, we derive an analytical closed-form expression for the end-to-end signal-to-noise plus interference ratio (SINR). Based on that, we derive the closed-form expressions for the outage probability and channel capacity of the considered UAV-based mmWave uplinks. The accuracy of the derived analytical expressions is verified by Monte Carlo simulations. Finally, we investigate the effects of different channel parameters such as antenna pattern gain, strength of UAV's vibrations, UAVs' positions in the sky, distribution of SBSs, and frequency reuse on the performance of the considered UAV-based uplink channel in terms of average capacity and outage probability.
\end{abstract}
\IEEEpeerreviewmaketitle

\section{Introduction}
\subsection{Background}
\IEEEPARstart{N}{ext}-\textcolor{black}{generation wireless networks are expected to connect a huge number of users/devices in ultra-dense networks with high data rate requirements for video streaming applications and ultra-low latency and high reliability in vehicle-to-vehicle communications \cite{giordani2020toward}. 
Small cells are considered a fundamental driver for the ongoing network densification \cite{8968715}. 
A small cell network consists of a series of small low-powered antennas that provide coverage and capacity in a similar way to a macrocell, with a few important distinctions \cite{8968715}.
%
%
%
With the dense deployment of small cell base stations (SBSs), fronthaul links demand a high capacity of more than 2.5 Gbps with a low latency of around 100 $\micro$s or less \cite{Ref20}. These demands can be traditionally fulfilled by coaxial cable or optical fiber in terrestrial networks. 
However, for massive deployment of small cells, such deployment will not be flexible, easy to deploy, and cost-effective as compared to wireless fronthaul links \cite{taori2015point}.
The wireless fronthaul connectivity can be realized using microwave bands for non-line-of-sight (NLoS) case or millimeter wave (mmWave) and free-space optical (FSO) links for line-of-sight (LoS) case \cite{9492795}. Microwave links can cover a wide area but suffer from low data rates. The FSO and mmWave based fronthaul links meet the capacity requirements of next generation communication networks, and are lightweight and easy to install. 
However, the mmWave/FSO links suffer from susceptibility to weather conditions \cite{alzenad2018fso} and require a LoS connection, which is the main hurdle in urban regions because small cells are mainly in hard-to-reach, near-street-level locations,different than macrocell BSs which are typically in more open, above-rooftop locations.
Recently, a scalable idea was presented in \cite{alzenad2018fso} that utilizes unmanned aerial vehicle (UAVs) as a wireless fronthaul hub point between small cells and the core network. These UAV-hubs acting as the networked flying platforms (NFPs) provide a possibility of wireless LoS fronthaul link and thus, overcome the limitations of few available wireless NLoS ground fronthaul links.
Uplink channel modeling of SBS to NFP fronthaul link is the main subject of this study which inevitably rely on two technologies, UAV and high-frequency mmWave bands and we hope that the results of this study will be useful for optimal design of future small cell backhaul/fronthaul links.}

\subsection{Literature Review and Motivation}
%
%
In order to use the benefits of employing UAV-assisted mmWave backhaul/fronthaul links for  SBSs, at first, it will be important to have an accurate and comprehensive channel model while taking into account the distribution of SBSs and NFP nodes as well as 3D mmWave directional antenna pattern and UAV's vibrations.
Although UAV-assisted channel modeling has been investigated in recent studies \cite{khuwaja2018survey,nguyen2018novel,matolak2017air,sun2017air}, these works are mainly limited to sub-6 GHz frequency bands which cannot be directly employed to UAV-based mmWave communication systems. Meanwhile, most of the prior studies on mmWave channel modeling \cite{rappaport2015wideband,ju2021millimeter,hemadeh2017millimeter} do not address the presence of UAVs, with the exception of a few recent works in \cite{gapeyenko2021line,          khawaja2017uav,khawaja2018temporal, 9426415,  
	    tafintsev2020aerial,kovalchukov2018analyzing,
	    rupasinghe2019non,rupasinghe2019angle,yapici2021physical,
	    gapeyenko2018flexible,
	kumar2020dynamic,dabiri2022enabling,dabiri2022study,dabiri2020analytical,dabiri20203d}.
For instance, a new LoS probability model for UAV to ground BS links is offered in \cite{gapeyenko2021line} over realistic urban grid deployments in which the effects of building height distributions along with their densities and dimensions are considered.
In \cite{khawaja2017uav,khawaja2018temporal}, the authors provided a new channel characterization for UAV-based mmWave links by using ray tracing simulations at two different frequency bands: 60 GHz and 28 GHz. 
In \cite{9426415}, a novel ground-to-UAV channel prediction method is provided by using ray tracing simulations based on the minimum Euclidean distance.
However, the results of these works are provided for omnidirectional mmWave antenna pattern.
Owing to the intrinsic feature of mmWave frequencies, nowadays, small size and light weight of high directional mmWave antennas are available in the market.
High directional antennas can adequately guide the signals towards desirable directions in order to efficiently increase security as well as to enhance the received signal power at an intended area and suppress interference at unintended areas \cite{lu2020robust}.
Furthermore, narrow directional beams can concentrate much more wireless energy on target users, to compensate for stronger propagation attenuation and higher free space path loss of mmWave frequencies. 
%
%

Based on the above-mentioned advantages, UAV-based directional mmWave link is the subject of recent works \cite{    tafintsev2020aerial,kovalchukov2018analyzing,
	rupasinghe2019non,rupasinghe2019angle,  yapici2021physical,
	gapeyenko2018flexible,
	kumar2020dynamic,dabiri2022enabling,dabiri2022study,dabiri2020analytical,dabiri20203d}
In order to have a better performance in a time varying traffic demand, a dynamic algorithm is offered in \cite{tafintsev2020aerial} to adjust UAV locations by considering clusterization and mobility of users, dynamic backhauling, and antenna array geometry. 
In \cite{kovalchukov2018analyzing}, the authors constructed a novel 3D model for UAV-based mmWave communication that consists the random heights of the communicating entities as well as the high directionality of transmissions.
The UAV-based scenarios are considered in \cite{rupasinghe2019non,rupasinghe2019angle} wherein non-orthogonal multiple access (NOMA) technique in mmWave frequencies is used to serve a  large number of mobile users simultaneously. Moreover, a NOMA-based transmission technique is employed in \cite{yapici2021physical} in order to increase the secrecy-rate performance of a UAV-based mmWave network in the presence of malicious devices.
In a different and new method, temporal and spatial characteristics of UAV-based mmWave backhaul performance are investigated in \cite{gapeyenko2018flexible} by taking into account the effects of 3D multi-path propagation along with heterogeneous mobility of users.
For UAV-based directional mmWave downlink communication scenario, a novel sectoring approach is presented in \cite{kumar2020dynamic} to ensure coverage of the whole area by taking into account the effects of interference caused by side lobe gain of antenna arrays.
However, the results of the aforementioned works are obtained for highly stable UAV with negligible UAVs' vibrations.
Directional mmWave communication links suffer from misalignment between receiver and transmitter. Due to the weight and power limitations for utilizing high quality stabilizers, perfect alignment is not practically feasible in aerial links, particularly, for small UAVs. This leads to an unreliable communication  due to antenna gain mismatch \cite{guan2019effects,zhong2019adaptive,pokorny2018concept}.

More recently, in \cite{dabiri2022study,dabiri2020analytical,dabiri20203d}, the authors investigated the relationship between UAV's vibrations and mmWave antenna gain. The results of these work clearly show that the performance of a UAV-based link with directional mmWave antenna is highly dependent on the strength of UAV's vibrations. 
However, the results of  \cite{dabiri2022study,dabiri2020analytical,dabiri20203d}  are provided for a special case without considering the effects of interference caused by side lobe gain of antenna array. In this study, we will show that in crowded networks, the effect of interference can not be ignored.

\subsection{Contributions and Paper Structure}

\begin{figure}
	\begin{center}
		\includegraphics[width=3.4 in]{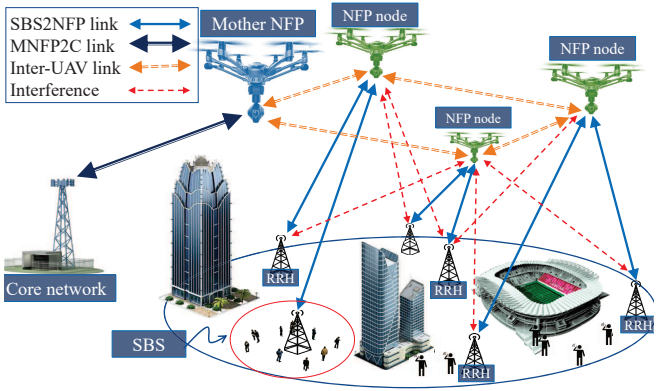}
		\caption{
			%
			An illustration of a UAV-assisted HetNet.
			As shown, the considered HetNet consists of three different wireless backhaul/fronthaul links: i) SBS-to-NFP links, ii) inter NFP links, and iii) mother NFP to core network  link. 
		}
		\label{rn1}
	\end{center}
\end{figure}

In this study, we consider a UAV-assisted heterogeneous network (HetNet) as shown in Fig. \ref{rn1} that is offered as a cost effective and easy to deploy solution in \cite{alzenad2018fso} to solve the problem related to transferring traffic of the distributed small cells to the core network.  
3D channel modeling of a SBS to aerial NFP uplink is the main contribution of this work by taking into account the realistic parameters. 
Our main contributions are summarized as follows:

\begin{itemize}
	\item \textcolor{black}{For any given distribution of SBSs, first, we characterize an accurate end-to-end signal-to-noise plus interference ratio (SINR) for SBS to aerial NFP by taking into consideration real parameters such as UAV's vibrations, distribution of SBSs, position of UAVs in the sky, real 3D antenna pattern model provided by 3GPP along with interference caused by antenna side lobes and frequency reuse.} 
	%
	\item For the characterized channel, we derive an analytical closed-form expression for the end-to-end SINR. We also derive the closed-form expressions for the outage probability and channel capacity of the considered UAV-based mmWave uplinks. 
	%
	%
	Then, by providing Monte Carlo simulations, the accuracy of the derived analytical expressions is verified.
	\item 
	We investigate the effects of key channel parameters such as antenna pattern gain, strength of UAV's vibrations, UAVs' positions in the sky, distribution of SBSs, and frequency reuse on the performance of the considered UAV-based uplink channel in terms of average capacity and outage probability.
	%
	By providing sufficient simulation results, we carefully study the relationships between these parameters in order to reduce interference and decrease outage probability and the same time, increase channel capacity as much as possible. 
\end{itemize}

The rest of this paper is organized as follows. We characterize the channel model of UAV-based mmWave network in Section II.
Then, in Section III, we provide the analytical channel model along with analytical closed-form expressions for outage probability and channel capacity.
Next, in Section IV, the performance of UAV-based mmWave network is analyzed. 
Finally, conclusions and future directions are drawn in Section V.

%

\begin{table}
	\caption{The list of main notations.} 
	\centering 
	\begin{tabular}{l l} 
		\hline\hline \\[-1.2ex]
		{\bf Parameter} & {\bf Description}  \\ [.5ex] 
		\hline\hline \\[-1.2ex]
		Subscript $i$ & Denote the $i$th sector \\
		Subscript $j$ & Denote the $j$th SBS in each sector \\
		%
		%
		$N_S$ & Total number of SBSs \\
		$N_D$ & Total number of UAVs acting as NFPs\\
		$A_s$ & Geographical area \\
		$A_{s_i}$ & Area of $i$th sector \\
		$U_i$ & Denote $i$th NFP where $i\in\{1,...,N_D\}$ \\
		$h_{U_i}$ & The height of $i$th NFP\\
		\textcolor{black}{$N_{SU_i}$} & \textcolor{black}{Number of array antennas mounted on $U_i$} \\
		$S_{i,j}$ & SBS associated to the $U_i$ for $j\in\{1,...,N_{SU_i}\}$\\
        %
        \textcolor{black}{$A_{r_{ij}}$s} & \textcolor{black}{Denote $j$th array antenna mounted on $U_i$}\\[-.7ex] 
          & \textcolor{black}{for $j=\{1,...,N_{{SU}_i}\}$}\\ [.5 ex]
        $P_{i,j}$ & Transmitted power by $S_{i,j}$ \\
        $L_{i,j}$ & Link length between $S_{i,j}$ and $U_i$ \\
        $L_{i',j}$ & Link length between $S_{i',j}$ and $U_i$ \\
        $h_L({L_x})$ & Path loss of a link with length $L_x$ \\
        $w_\textrm{ma}$ & The maximum bandwidth allocated to each NFP \\
        $w_{i,j}$ & Frequency band dedicated to the $S_{i,j}$ to $U_i$ link \\
        $R_u$ & Frequency reuse number for each sector \\
        $N_{{i,j}}$ & Denote  square array antenna of $S_{i,j}$ with \\[-.7ex]
        &$N_{i,j}\times N_{i,j}$ elements\\ [.5ex]
        $N'_{i,1}$ & Denote square antenna of $A_{r_{i,1}}$ with \\[-.7ex]
        &$N'_{i,1}\times N'_{i,1}$ elements\\[.5ex]
		$[x; y; z]$ & Cartesian coordinate system that  $z$ axis refers to \\[-.7ex]
		& the direction that extends from $A_{r_{i,1}}$ toward $S_{i,1}$ \\ [.5ex]
		$\lambda$ and $f_c$ &  Wavelength and carrier frequency \\
		$\alpha$ nad $\beta$ & constants whose values depend on  the\\[-.7ex]
		 &  propagation environment\\[.5ex]
		$\Gamma_{i,1}$ & SINR of considered $S_{i,1}$ to $U_i$ uplink \\
		$\sigma_n^2$ &  The thermal noise power \\ 
		%
		%
		%
		\hline
		$\theta_{xU_{i,1}}$  & Instantaneous orientation of $A_{r_{i,1}}$ in the $x-z$ \\[-.7ex] 
		&  Cartesian coordinates (CC) \\[.5ex]
		$\theta_{yU_{i,1}}$  & Instantaneous orientation of $A_{r_{i,1}}$ in $y-z$ CC \\ 
		$\theta_{xU_{i,j}}$ & Direction angle (DA) from $U_i$ to $S_{i,j}$   \\ [-.7ex] 
	                      	& in $x-z$ CC \\ [.5ex]
		$\theta_{yU_{i,j}}$ & DA from $U_i$ to $S_{i,j}$ in $y-z$ CC\\
		$\theta_{xU_{i',j}}$ & DA from $U_i$ to $S_{i',j}$ in $x-z$ CC \\
		$\theta_{yU_{i',j}}$ & DA from $U_i$ to $S_{i',j}$ in $y-z$ CC\\
		$\theta'_{xU_{i',j}}$ & AoD of $S_{i',j}$ toward $U_i$ in $x-z$ CC  \\
		$\theta'_{yU_{i',j}}$ & AoD of $S_{i',j}$ toward $U_i$ in $y-z$ CC\\
	    $\theta_{U_{i,1}}$    &  $   = \tan^{-1}\left(\sqrt{\tan^2(\theta_{xU_{i,1}})+\tan^2(\theta_{xU_{i,1}})}\right)$ \\
		$\psi_{U_{i,1}}$      &  $   = \tan^{-1}\left(\frac{\tan(\theta_{yU_{i,1}})}{\tan(\theta_{xU_{i,1}})}\right)$ \\
		$\theta_{\textrm{elev}i',j}$ & Elevation angle of $S_{i',j}$ compared to $U_i$\\
		%
		%
		\hline
		$\mu_x$ \& $\mu_y$ & Mean of RVs $\theta_{xU_{i,1}}$ and $\theta_{yU_{i,1}}$\\
		$\sigma_\theta^2$ & Variance of RVs $\theta_{xU_{i,1}}$ and $\theta_{yU_{i,1}}$\\
		\hline
		$G_{U_{i,1}}$  & The received antenna pattern gain   \\[-.7ex]
		& of $A_{r_{i,1}}$ relative to $S_{i,1}$\\[.5ex]
		$G_{S_{i,1}}$  &The transmitted antenna pattern gain   \\ [-.7ex] 
		&of $S_{i,1}$ directed toward $U_i$ \\  [.5ex]
		\hline 
		$f_x(x)$ & The PDF of RV $x$ \\
		$F_x(x)$ & The CDF of RV $x$ \\
		%
		%
		%
		%
		\hline \hline              
	\end{tabular}
	\label{I1} 
\end{table}

\section{System Model}
%
We consider a UAV-assisted HetNet as shown in Fig. \ref{rn1} that is offered as a cost effective and easy to deploy solution in \cite{alzenad2018fso} to solve the problem related to transferring traffic of the distributed small cells to the core network. More precisely, as an alternative for backhaul/fronthaul links, the UAV-based mmWave or FSO communication links are proposed in \cite{alzenad2018fso} to carry the small cell traffic, particularly in ultra-dense urban areas. 
However, under foggy, raining and cloudy conditions, the FSO communication link fails because the received power is less than the sensitivity of the receiver \cite{alzenad2018fso}. Unlike FSO communications, mmWave are not attenuated by fog. Therefore, in our system model, we use mmWave for backhaul/fronthaul links of the considered system. 
%
\textcolor{black}{As we show in Fig. \ref{rn1}, the considered HetNet consists of three different wireless backhaul/fronthaul links: i) SBS-to-NFP (SBS2NFP) links, ii) inter NFP links, and iii) mother NFP to core network (MNFP2C) link. In this paper, our aim is to model SBS2NFP link which is a fronthaul link that connects the  remote radio head (RRH) to the NFP.}
Moreover, for a SBS2NFP fronthaul link, a downlink is the link from a NFP down to a SBS, and an uplink is the link from a SBS up to a NFP.
As we show next, many parameters affect the performance of the uplink and the downlink, and analysis of all parameters on the performance of these links requires enough space. Due to space constraints, in this work, we focus on uplink channel modeling and the channel modeling of downlink can be the subject of another work.

%
\begin{figure}
	\begin{center}
		\includegraphics[width=3.2 in]{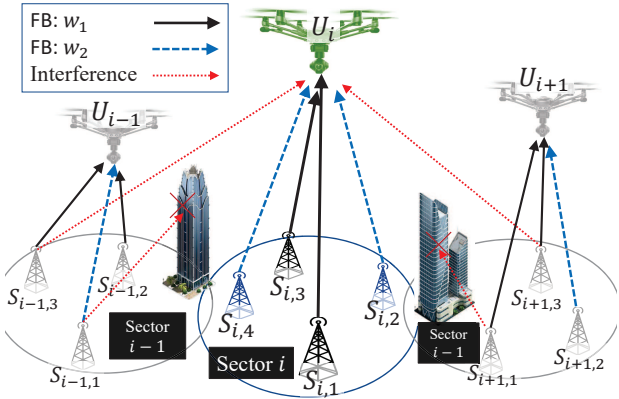}
		\caption{A graphical illustration of SBSs to NFPs uplinks. In this topology, a set including $S_{i,j}$s which are connected to $U_i$ is called the $i$th sector. As a graphical example, the sector $i$ consists four SBSs that both $S_{i,1}$ and $S_{i,3}$ are connected to $U_i$ with the same frequency band $w_1$ and both $S_{i,2}$ and $S_{i,4}$ are connected to $U_i$ with the same frequency band $w_2$. For instance, $S_{i,3}$ to $U_i$ uplink causes an intra-sector interference on $S_{i,1}$ to $U_i$ uplink. Also, $S_{i+1,3}$ causes an inter-sector interference on $S_{i,1}$ to $U_i$ uplink. However, the inter-sector interference causes by $S_{i+1,1}$ is blocked by a tall building.}
		\label{rn2}
	\end{center}
\end{figure}
%

\subsection{3D Antenna Pattern}
%
Due to the UAVs' transmission power constraints, using antennas with high gain is needed to combat severe propagation loss, particularly for high data rate and longer backhaul links. Advances in the fabrication of antenna array technology at mmWave bands allow the creation of large antenna arrays with high gain in a cost effective and compact form. For instance, light-weight directional mmWave array antennas are already available in the market, which are suitable to be mounted on UAVs with limited payload. In addition, and as we will show, employing directional mmWave antenna pattern allows us to reuse frequency bands and thus, it increases the spectral efficiency of the considered system which is very important for 5G+ system. 

Due to an approximate symmetry in the UAV vibrations in the $x$- and $y$-directions, we consider a uniform square array antenna, comprising $N\times N$ antenna elements with the same spacing between elements in $x$- and $y$-directions, i.e., $d_{x}=d_{y}=d_a$ where $d_{x}$ and $d_{y}$ are the spacing between antenna elements in $x$- and $y$-direction, respectively.
The array radiation gain is mainly formulated in the direction of $\theta$ and $\phi$. 
By taking into account the effect of all elements, the array radiation gain in the direction of angles $\theta_{x}$ and $\theta_{y}$ will be:
\begin{align}
	\label{p_1}
	G(\theta,\phi)  = G_0(N) \,
	\underbrace{G_e(\theta,\phi) \,  G_a(\theta,\phi)}_{G'(\theta,\phi)},
\end{align}
where $G_a$ is an array factor, $G_e$ is single element radiation pattern and $G_0$ is a constant defined in the sequel. 
From the 3GPP single element radiation pattern, $G_{e,\textrm{3dB}}=10\times\log_{10}(G_e)$ of each single antenna element is obtained as \cite{niu2015survey}
\begin{align}
	\label{vb1}
	&G_{e\textrm{3dB}} = G_{\textrm{max}} - \min\left\{-(G_{e\textrm{3dB,1}}+G_{e\textrm{3dB,2}}),F_m 
	\right\}, \nonumber \\
	&G_{e\textrm{3dB,1}} =  -\min \left\{ - 12\left(\frac{\theta_e-90}{\theta_{e\textrm{3dB}}}\right)^2,
	G_{\textrm{SL}}\right\},      
	\nonumber \\
	&G_{e\textrm{3dB,2}} = -\min \left\{ - 12\left(\frac{\theta_{x}}{\phi_{e\textrm{3dB}}}\right)^2,
	F_m\right\},
	\nonumber \\
	&\theta_e             = \tan^{-1}\left( \frac{\sqrt{1+\sin^2(\theta_{x})}}
	{\sin(\theta_{y'})} \right), 
\end{align}
where $\theta_{e\textrm{3dB}}=65^{\circ}$ and $\phi_{e\textrm{3dB}}=65^{\circ}$ are the vertical and horizontal 3D beamwidths, respectively, $G_{\textrm{max}}=8$ dBi is the maximum directional gain of the antenna element, $F_m=30$ dB is the front-back ratio, and $G_{\textrm{SL}}=30$ dB is the side-lobe level limit. 

If the amplitude excitation of the entire array is uniform, then the array factor $G_a(\theta,\phi)$ for a square array of $N\times N$ elements can be obtained as \cite[eqs. (6.89) and (6.91)]{balanis2016antenna}
\begin{align}
	\label{f_1}
	G_a(\theta, \phi) &= 
	\left( \frac{\sin\left(\frac{N (k d_{x} \sin(\theta)\cos(\phi)+\beta_{x})}{2}\right)} 
	{N\sin\left(\frac{k d_{x} \sin(\theta)\cos(\phi)+\beta_{x}}{2}\right)}
	\right. \nonumber \\
	&\times \left. \frac{\sin\left(\frac{N (k d_{y} \sin(\theta)\sin(\phi)+\beta_{y})}{2}\right)} 
	{N\sin\left(\frac{k d_{y} \sin(\theta)\sin(\phi)+\beta_{y}}{2}\right)}\right)^2,
\end{align}
where $d_{x}=\frac{\lambda}{2}$ and $\beta_{x}$ are the spacing and progressive phase shift between the elements along the  $x$ axis, respectively, $d_{y}=\frac{\lambda}{2}$ and $\beta_{y}$ are the spacing and progressive phase shift between the elements along the $y$ axis, respectively, $k=\frac{2\pi}{\lambda}$ denotes the wave number, $\lambda=\frac{c}{f_c}$ denotes the wavelength, $f_c$ denotes the carrier frequency and $c$ is the speed of light. 

One of our key goals is to answer the question: for a UAV with a given instability as well as a given network topology, what is the optimum values of $N$ that achieves maximum throughput?
For a fair comparison between antennas with different $N$, we consider the total radiated power of antennas with different $N$ are the same. From this, we have
\begin{align}
	\label{cv}
	G_0(N)=\frac{1}{\int_0^{\pi}\int_0^{2\pi} G'(\theta,\phi) \sin(\theta) d\theta d\phi}.
\end{align}
More details on the element and array radiation pattern is provided in \cite{niu2015survey,balanis2016antenna}.
In addition, and without loss of generality, it is assumed that $\beta_{x}=\beta_{y}=0$ and the hovering UAV sets its antenna main-lobe direction on the $z$ axis. 
%

\begin{figure}
	\centering
	\subfloat[] {\includegraphics[width=2.2 in]{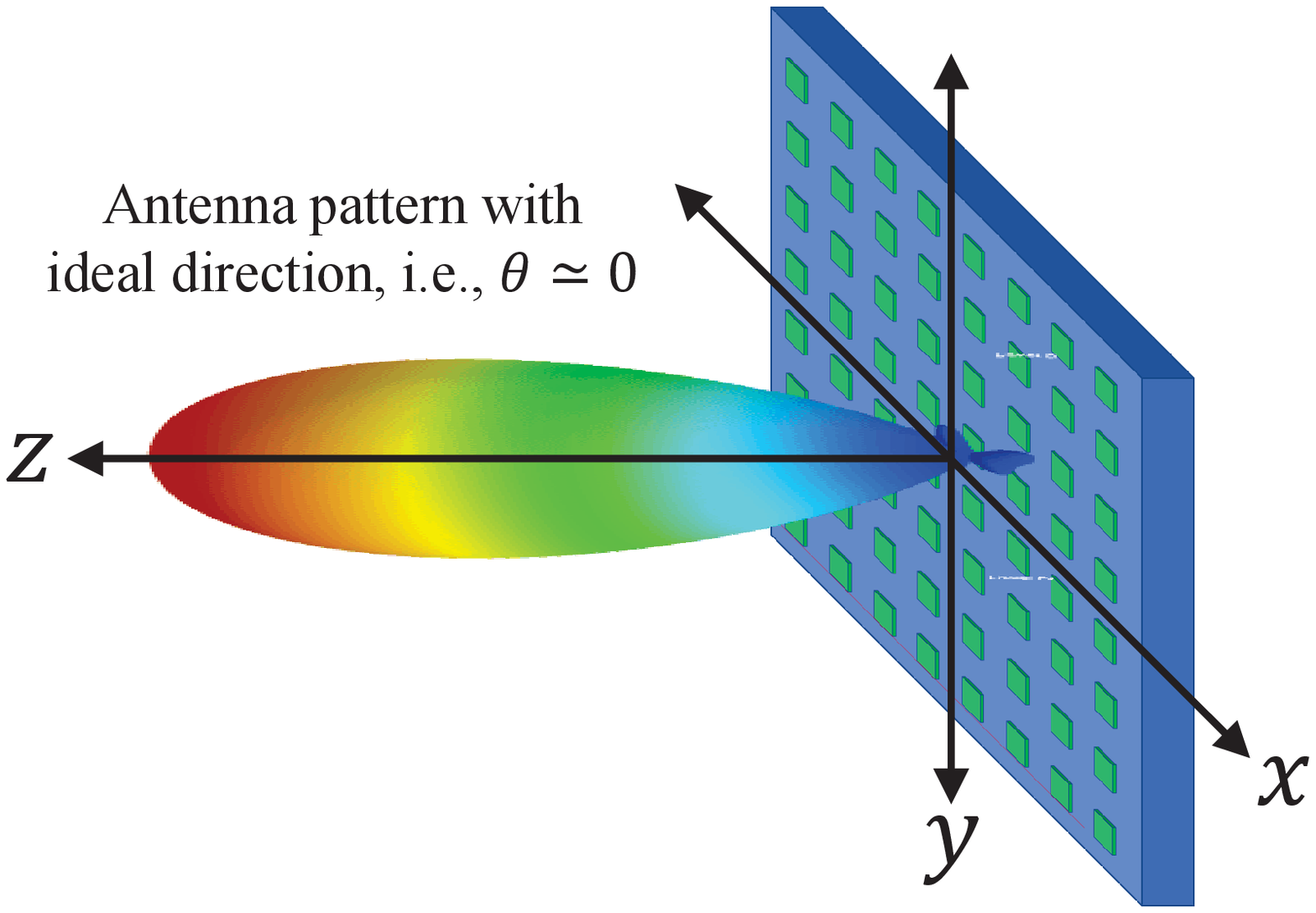}
		\label{xv1}
	}
	\hfill
	\subfloat[] {\includegraphics[width=3.35 in]{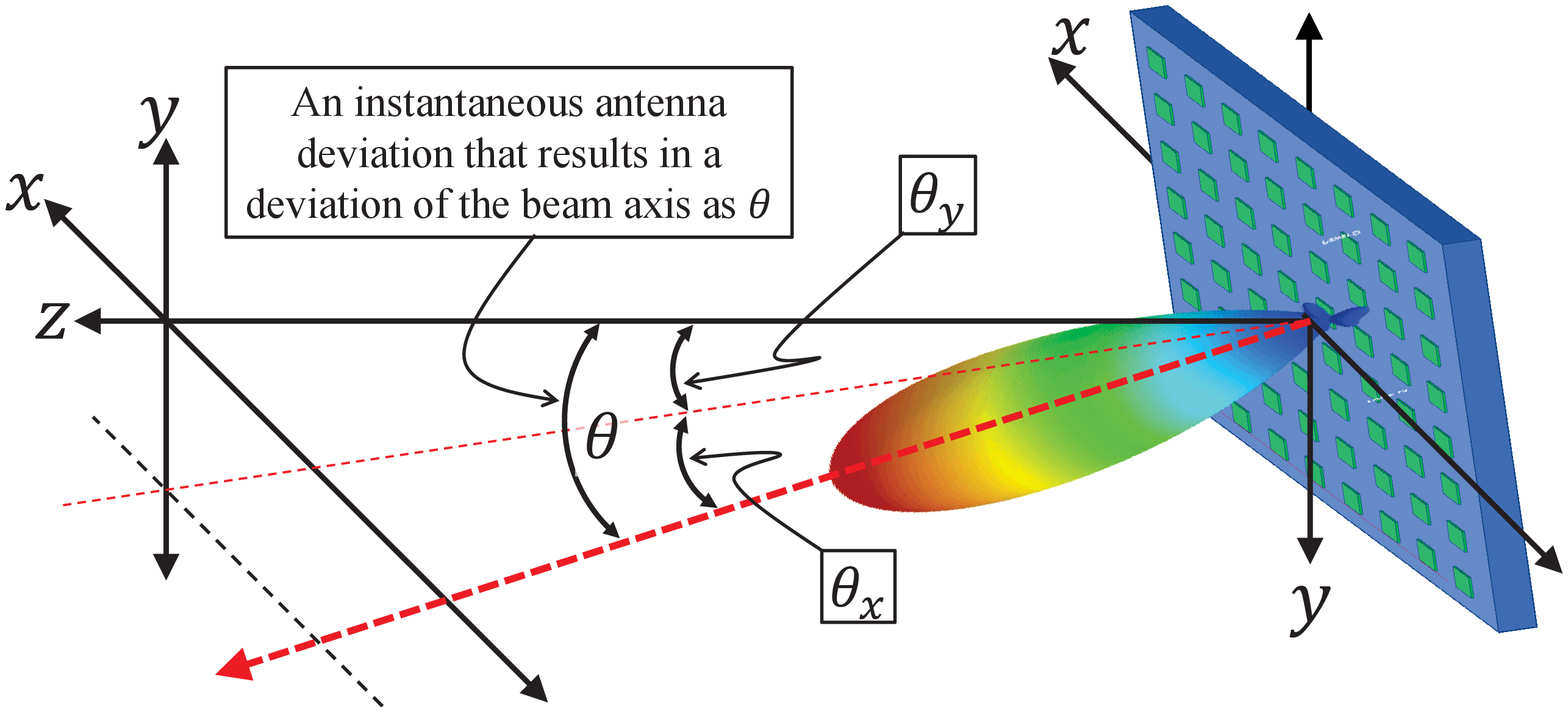}
		\label{xv2}
	}
	\caption{\textcolor{black}{3D illustration of an antenna pattern generated by a uniform $N\times N$ antenna array: (a) showing the ideal condition of a $N\times N$ square array antenna arranged in $x-y$ plane with $\theta\simeq0$; (b) showing the instantaneous orientation of UAV that leads to deviations in AoA of antenna pattern mounted on the UAV. The antenna orientation fluctuations are denoted by $\theta_{x}$ and $\theta_{y}$ in the $x-z$ and $y-z$ Cartesian coordinates, respectively.}}
	\label{xv4}
\end{figure}

\textcolor{black}{
	\subsection{Topology Description}
	For our topology, we consider $N_{S}$ SBSs and $N_D$ UAVs (acting as NFPs) which are distributed randomly over a geographical area of $A_s$. Let us denote each NFP with $U_i$ for $i\in\{1,...,N_D\}$.
	Also, we consider $U_i$ is equipped by $N_{{SU}_i}$ directional antenna denoted by $A_{r_{ij}}$s for $j\in\{1,...,N_{{SU}_i}\}$.\footnote{\textcolor{black}{
			In this study, we consider simple square array antennas without the use of any electrical or mechanical phase shifter that the power associated to each antenna elements is the same. 
			Note that in this simple structure, for the considered frequencies above 60 GHz, the size of the array antennas is in the order of several tens of millimeters, which is smaller, lighter and most importantly much cheaper than the phase array antennas \cite{
				ghattas2020compact,asaadi2018high,saily2011low}. 
			For instance, the size of the manufactured array antenna provided in \cite{saily2011low} at 60 GHz with a maximum gain of 21 dBi including the reactive power dividers is $35\times 35$ mm$^2$.}} 
	In other words, each $U_i$ can be connected to $N_{SU_i}$ SBSs where we must have $\sum_{i=1}^{N_D} N_{SU_i}= N_S$.
	%
	%
	Also, each SBS associated to the $U_i$ is denoted by $S_{i,j}$ where $j\in\{1,...,N_{SU_i}\}$.
	We denote the antenna of each $S_{ij}$ by $A_{t_{ij}}$ that the direction of $A_{t_{ij}}$ is adjusted toward the $U_i$.
	During employing directional antenna, it is essential that the main lobes of the transmitter and receiver antennas be aligned. 
	Due to the weight and power limitations of NFPs for utilizing high quality stabilizers, it is more practical to use simple stepper motors to align the antennas. The use of a simple stepper motor can only adjust the average direction of $A_{r_{ij}}$ towards $A_{t_{ij}}$, in which case it is expected that the instantaneous NFP's vibrations directly cause the instantaneous misalignment between the antennas which is modeled in the next subsection.
	Another point is that the dimensions of the considered array antennas without using any electrical or mechanical phase shifter at frequencies higher than 60 GHz is in the order of 
    few tens of millimeters \cite{ghattas2020compact,asaadi2018high,saily2011low}, and thus, it is possible to adjust the average direction of $A_{r_{ij}}$s by using smaller motors with lower power consumption.}

\subsection{The Effect of UAV's Instabilities}
Based on the above results, at first, it may seem that by increasing the antenna gain, the performance of the considered small cell to NFP node increases and at the same time the interference decreases. However, this result can be true for the ideal state, i.e., a stable NFP node without any orientation and position fluctuations. \textcolor{black}{In practical situations, an error in the mechanical control system of UAVs, mechanical noise, position estimation errors, air pressure, and wind speed can affect the UAV's angular and position stability.
In practice, the instantaneous orientation of a UAV can randomly deviate from its means denoted by $\theta$. This, in turn, leads to deviations in the AoD of Tx and/or AoA of Rx antenna pattern.
As shown in Fig. \ref{xv2}, the antenna orientation fluctuations of antennas mounted on the UAV are denoted by $\theta_{x}$ and $\theta_{y}$ in the $x-z$ and $y-z$ Cartesian coordinates, respectively. 
Based on the central limit theorem, the UAV's orientation fluctuations are considered to be Gaussian distributed \cite{dabiri2018channel,kaadan2014multielement,dabiri2019tractable}. 
Therefore, we have 
$\theta_{x}\sim \mathcal{N}(\mu_x,\sigma^2_{\theta})$, and $\theta_{y}\sim \mathcal{N}(\mu_y,\sigma^2_{\theta})$.
In our model, as graphically illustrated in Fig. \ref{xv2}, the RVs $\theta$ and $\phi$ can be defined as  functions of random variables (RVs) $\theta_{x}$ and $\theta_{y}$ as follows:
\begin{align}
	\label{f_2}
	\theta  &= \tan^{-1}\left(\sqrt{\tan^2(\theta_{x})+\tan^2(\theta_{y})}\right), \nonumber\\
	\phi    &=\tan^{-1}\left(\frac{\tan(\theta_{y})}{\tan(\theta_{x})}\right).
\end{align}}

\textcolor{black}{It can be easily shown that by increasing the antenna pattern gain, the beamwidth decreases and thus, the performance of the considered UAV-based system becomes more sensitive to the UAV's fluctuations.}
Moreover, the UAV’s fluctuations can have a significant effect on interference caused by the side lobes. Therefore, for any given strength of UAVs’ vibrations, optimizing radiation pattern shape requires balancing an inherent tradeoff between decreasing pattern gain to alleviate the adverse effect of a UAV’s vibrations and increasing it to compensate the undesired interference along with the large path loss at mmWave frequencies.

\textcolor{black}{
\subsection{SINR Definition}
In this paper, as shown in Fig. \ref{rn2}, the set including $U_i$ and $S_{ij}$s is called the $i$th sector. 
For the considered system, there are two different types of interference: i) intra-sector interference, and ii) inter-sector interference. 
Without loss of generality and for notation simplicity, in the sequel we model the uplink between $S_{i,1}$ to $A_{r_{i,1}}$.
Therefore, the SINR of the considered uplink $S_{i,1}$ to $U_i$ is formulated as 
\begin{align}
		\label{po1}
		\Gamma_{i,1} = \frac{\mathbb{R}_{i,1}}  { \mathbb{I}_{\text{intra}}  + 
			\mathbb{I}_{\text{inter}}  +  \sigma_n^2 },
\end{align}
where $\sigma_n^2$ is the thermal noise power, $\mathbb{I}_{\text{inter}}$ and $\mathbb{I}_{\text{intra}}$ are respectively inter- and intra-sector interference that will be exactly modeled in the next section.}

\subsection{Frequency Allocation}
The maximum bandwidth allocated to each NFP is $w_\textrm{ma}$. At first, it may seem that to reduce the interference, it is better to allocate a separate band to each $S_{ij}$ to $U_i$ link. In this case, regardless of the guard bands, the bandwidth allocated to each $S_{ij}$ to $U_i$ link is approximately equal to $w_{ij}=\frac{w_\textrm{ma}}{N_{{SU}_i}}$. However, this causes a waste of bandwidth. 
In this paper, by using directional antenna pattern, we will show how several SBSs can be connected to a NFP with the same frequency band (FB) at the same time. In this case, the bandwidth allocated to each $S_{ij}$ to $U_i$ link increases as
\begin{align}
	w_{ij}=\frac{w_\textrm{ma}}{N_{{SU}_i}}\times R_u,
\end{align} 
where $R_u$ is the frequency reuse for the links connected to a NFP. 
For example, Fig. \ref{rn2} is drawn for the $R_u=2$ and $N_{{SU}_i}=4$ case. As shown, the SBSs $S_{i,1}$ and $S_{i,3}$ are connected to $U_i$ by the same FB $w_1$, i.e., $w_{i,1}=w_{i,3}=w_1$. Also, the SBSs $S_{i,2}$ and $S_{i,4}$ are connected to $U_i$ by the same FB $w_2$, i.e., $w_{i,2}=w_{i,4}=w_2$.
Now, the instantaneous channel capacity of the considered $S_{i,j}$ to $U_i$ uplink is formulated as
\begin{align}
	\label{gt1}
	\mathbb{C}_i = \frac{w_\textrm{ma}}{N_{{SU}_i}}\times R_u \log\left(1+ \Gamma_{i,1}\right),
\end{align}
where $\Gamma_{i,1}$is the instantaneous SINR of $S_{i,j}$ to $U_i$ uplink and will be formulated in the next section.
\subsection{Propagation Channel Loss}
Since there is still no standardized results for UAV-based communications at mmWave bands, we consider the results of the recent 3GPP report in \cite{3gppf} in order to set the path loss parameters. These parameters are valid for a BS height up to  150 m and are expressed as follows
\begin{align}
	\label{loss}
	h_{\textrm{L,dB}}(L) &= -20 \log_{10}\left(\frac{40 \pi L f_c}{3}\right) \\ 
	&~~~+ \min\left\{0.03 h_b ^{1.73},10\right\}\times \log_{10}(L) \nonumber \\
	&~~~ + \min\left\{0.044 h_b ^{1.73},14.77\right\}
	-0.002\, \times L \log_{10}(h_b),  \nonumber
\end{align}
where  $L$ is the link length and $h_b$ (in meter) is the average of building height of the city.

In addition to the high propagation loss, mmWave communication systems are very sensitive to blockages \cite{bai2014coverage}. Even the human body can reduce the signal strength by 20 dB \cite{maccartney2017rapid} and other signal blockages such as walls and buildings reduce signal strength much more than 20 dB. Backhaul links using the mmWave band (V-band or the E-band) are well suited to supporting 5G due to their 10 Gbps to 25 Gbps data throughput capabilities. Such a data rate requires a high SINR at the receiver and thus, in the presence of physical objects such as walls and buildings, the SINR at the receiver will be below the threshold level and the communication link will be in outage.
Therefore, the probability of LoS is an important factor and can be described as a function of the elevation angle and environment as follows \cite{al2014modeling,al2014optimal}:
\begin{align}
\label{op11}
P_\textrm{LoS}(\textrm{elev}) = \frac{1}
{1+\alpha \exp\left(-\beta(\frac{180}{\pi}\theta_\textrm{elev}-\alpha)\right)}	
\end{align}
where $\alpha$ and $\beta$ are constants whose values depend on the propagation environment, e.g., rural, urban, or dense urban, and $\theta_\textrm{elev}$ is the elevation angle.

\section{Performance Analysis}

%
\begin{figure}
	\begin{center}
		\includegraphics[width=3.4 in]{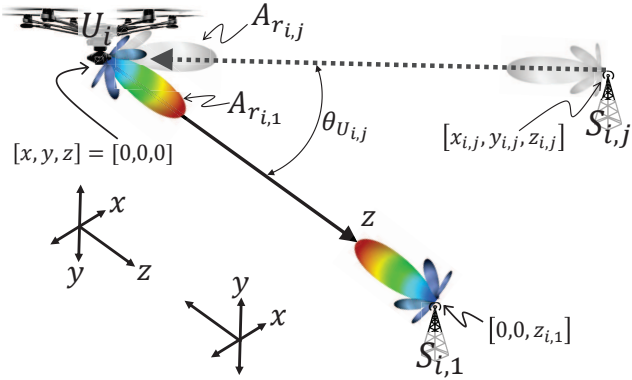}
		\caption{An illustration of how intra-sector interference occurs for the $i$th sector with $R_u=2$ wherein the SBSs $S_{i,1}$ and $S_{i,j}$ are  respectively connected  to antennas $A_{r_{i,1}}$ and $A_{r_{i,j}}$ (which are mounted on $U_i$) with the same frequency band.
	    As shown, $z$ axis refers to the direction that extends from $A_{r_{i,1}}$ toward $S_{i,1}$.}
		\label{rb1}
	\end{center}
\end{figure}
%

%
\begin{figure}
	\begin{center}
		\includegraphics[width=3.2 in]{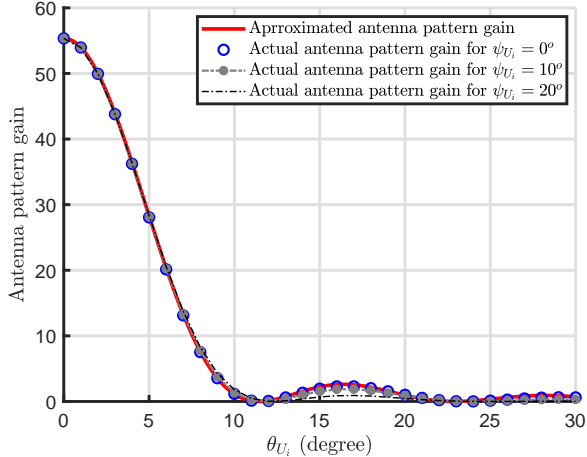}
		\caption{\textcolor{black}{Actual antenna pattern obtained by \eqref{p_1} versus $\theta_{U_{i,1}}$ for different values of $\psi_{U_{i,1}}$ and comparison with 
			approximated pattern used in \eqref{xvs}.}}
		\label{rb3}
	\end{center}
\end{figure}
%

\begin{figure*}
	\centering
	\subfloat[] {\includegraphics[width=3.5 in]{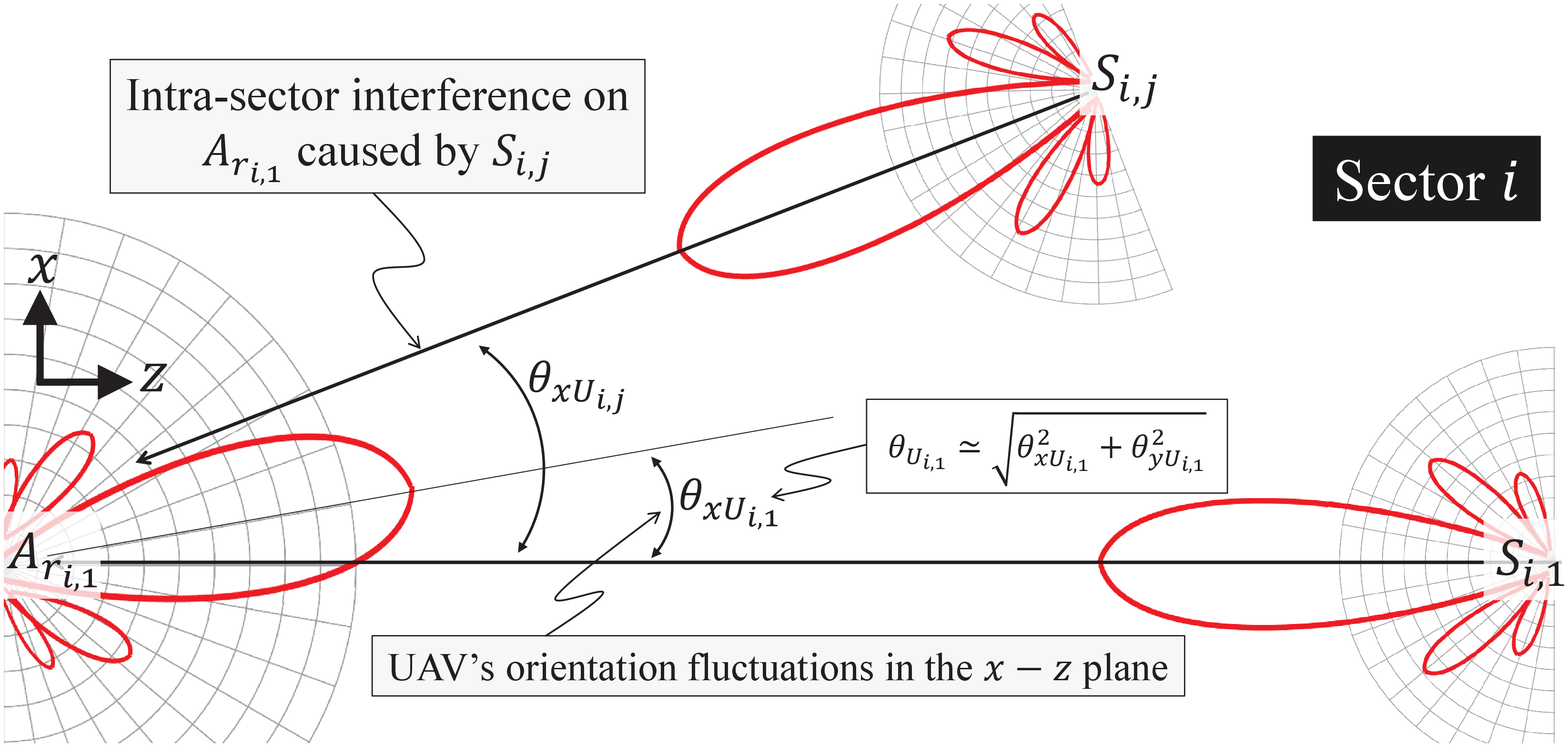}
		\label{nv1}
	}
	\hfill
	\subfloat[] {\includegraphics[width=3.5 in]{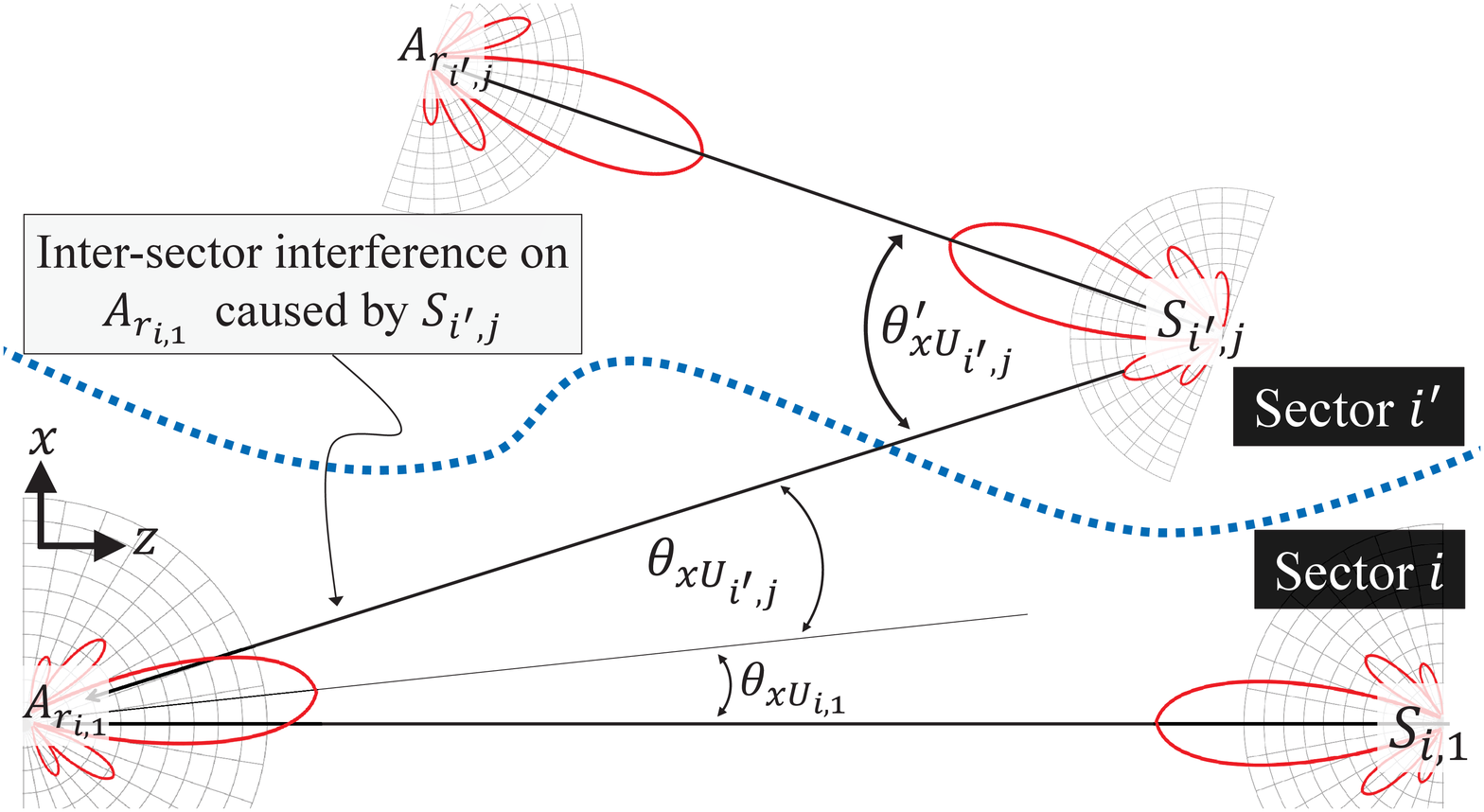}
		\label{nv2}
	}
	\caption{Graphical illustration of interference in 2D $x-z$ plane: (a) intra-sector interference caused by main lobe of $S_{i,j}$ on $A_{r_{i,1}}$, and (b) inter-sector interference caused by the side lobes of $S_{i',j}$. 
	The hovering $U_i$ node tries to set the main-lobe of $A_{r_{i,1}}$ in the direction of $z$-axis, however, the instantaneous orientation of the antenna of $A_{r_{i,1}}$ can randomly deviate from its means denoted by $\theta_{U_{i,1}}$.
}
	\label{nv4}
\end{figure*}

In this section, we first characterize the end-to-end SINR of the uplink, and then we provide closed-form expressions for the PDF of the uplink as well as for channel capacity and outage probability.

\subsection{Received Signal Transmitted by $S_{i,1}$}
As defined in Section II, $A_{r_{i,1}}$ is one of the $N_{{SU}_i}$ antennas mounted on $U_i$ that its main lobe is adjusted toward $S_{i,1}$ by using a simple step motor.
The transmitted signal by $S_{i,1}$ collected by $A_{r_{i,1}}$ is modeled as
\begin{align}
	\label{sd01}
	\mathbb{R}_{i,1} = P_{i,1} h_L({L_{i,1}})
	\mathbb{G}_{i,1}(\theta_{U_{i,1}},\psi_{U_{i,1}} ,  \theta_{S_{i,1}},\psi_{S_{i,1}} ),
\end{align}
where
\begin{align}
	\label{sd02}
	&\mathbb{G}_{i,1}(\theta_{U_{i,1}},\psi_{U_{i,1}} ,  \theta_{S_{i,1}},\psi_{S_{i,1}} ) = \\
	&~~~~~~~~~~~~~~~ G_{U_{i,1}}     (\theta_{U_{i,1}},\psi_{U_{i,1}})
	\times G_{S_{i,1}} (\theta_{S_{i,1}},\psi_{S_{i,1}}), \nonumber
\end{align}
and $G_{U_{i,1}}   (\theta_{U_{i,1}},\psi_{U_{i,1}})$ is the received antenna pattern gain of $A_{r_{i,1}}$ relative to $S_{i,1}$ and 
$G_{S_{i,1}}   (\theta_{S_{i,1}},\psi_{S_{i,1}})$ is the transmitted antenna pattern gain of $S_{i,1}$ directed toward $U_i$.
%
%
We consider the positions of $U_i$ and $S_{i,1}$ are located at $[0,0,0]$ and $[0,0,z_i]$ in Cartesian coordinate system $[x,y,z]\in\mathcal{R}^3$, respectively.
As shown in Fig. \ref{rb1}, $z$ axis refers to the direction that extends from $A_{r_{i,1}}$ toward $S_{i,1}$.
The hovering $U_i$ node sets the main-lobe of $A_{r_{i,1}}$ in the direction of the $z$-axis. In practice, the instantaneous orientation of the antenna mounted on UAV $A_{r_{i,1}}$ can randomly deviate from its means denoted by $\theta_{U_{i,1}}$ and $\phi_{U_{i,1}}$ as depicted in Fig. \ref{nv1}.
The SBSs are stable compared to the UAV node acting as NFP. Also, the SBSs do not face weight and power limitations, so it is possible for the SBSs to use a stabilizer to align their antennas toward the UAV node.
Therefore, unlike the unstable UAV node, it is a practical assumption that $S_{i,j}$s are perfectly aligned to aerial $U_i$ node.
From this point, the random variables $\theta_{S_{i,1}}$ and $\psi_{S_{i,1}}$ tend to zero and thus, \eqref{sd02} can be simplified as 
\begin{align}
	\label{sd03}
	\mathbb{G}_{i,1}(\theta_{U_{i,1}},\psi_{U_{i,1}} ) =  G_{\textrm{max}S_{i,1}} G_{U_{i,1}}     (\theta_{U_{i,1}},\psi_{U_{i,1}})  
\end{align}
where $G_{\textrm{max}S_{i,1}} = G_{S_{i,1}} (\theta_{S_{i,1}}=0,\psi_{S_{i,1}}=0)$.
The RVs $\theta_{U_{i,1}}   = \tan^{-1}\left(\sqrt{\tan^2(\theta_{xU_{i,1}})+\tan^2(\theta_{xU_{i,1}})}\right)$ and 
$\psi_{U_{i,1}}     = \tan^{-1}\left(\frac{\tan(\theta_{yU_{i,1}})}{\tan(\theta_{xU_{i,1}})}\right)$ are functions of RVs 
$\theta_{xU_{i,1}}\sim \mathcal{N}(0,\sigma^2_{i})$ and 
$\theta_{yU_{i,1}}\sim \mathcal{N}(0,\sigma^2_{i})$.
%
%
As shown in Fig. \ref{xv2}, the parameters $\theta_{xU_{i,1}}$ and $\theta_{yU_{i,1}}$ are the antenna orientation fluctuations in the $x-z$ and $y-z$ Cartesian coordinates, respectively.

To easily simulate and also derive a tractable analytical model for \eqref{po1}, we must first calculate the analytical model for 
\eqref{sd03}. As we observe from \eqref{p_1}, \eqref{vb1} and \eqref{f_1}, the array antenna gain is a complex function of $\theta_{xU_{i,1}}$ and $\theta_{yU_{i,1}}$. 
For low values of $x$ and $A\,x$, the expression $\sin(A\sin(x))$ can be approximated as $ A\,x$. Using this, we approximate \eqref{sd03} as 
\begin{align}
	\label{xvs}
	&\mathbb{G}_{i,1}(\theta_{xU_{i,1}},\theta_{yU_{i,1}} ) \simeq   
	G_n G_{\textrm{max}S_{i,1}} \\
	&\times \frac{1-\cos\left( N'_{i,1} k d_a  \tan^{-1}\left(\sqrt{\tan^2(\theta_{xU_{i,1}})+\tan^2(\theta_{yU_{i,1}})}\right)\right)}
	{  {N'_{i,1}}^2 \left( \tan^{-1}\left(\sqrt{\tan^2(\theta_{xU_{i,1}})+\tan^2(\theta_{yU_{i,1}})}\right)\right)^2}, \nonumber
\end{align}
where $G_n = 3.1548$. 
In Fig. \ref{rb3}, the approximated pattern generated by \eqref{xvs} is compared with the actual antenna pattern obtained by \eqref{p_1} versus $\theta_{U_{i,1}}$ for different values of $\psi_{U_{i,1}}$.
As we observe, an exact match exists between the approximated model and the actual antenna pattern, specially, at the main-lobe.

\subsection{Intra-Sector Interference Analysis}
As mentioned in the previous section, in order to increase the network capacity, it is possible to reuse the spectrum in each sector by using the directional antennas.
However, reusing the frequency causes intra-sector interference. To find a better view of the intra-sector interference, as an example, Fig. \ref{rb1} is drawn for the $i$th sector with $R_u=2$. As depicted in Fig. \ref{rb1}, SBSs $S_{i,1}$ and $S_{i,j}$ are  respectively connected  to antennas $A_{r_{i,1}}$ and $A_{r_{i,j}}$ (which are mounted on $U_i$) with the same frequency band. 
The directions of $A_{r_{i,1}}$ and $S_{i,1}$ antennas are set towards each other and the directions of $A_{r_{i,j}}$ and $S_{i,j}$ antennas are set towards each other. 
However, both uplinks $A_{r_{i,1}}$ to $S_{i,1}$ and $A_{r_{i,j}}$ to $S_{i,j}$ cause an intra-sector interference on each other.
The intra-sector interference caused by $S_{i,j}$ on $A_{r_{i,1}}$ is graphically depicted in Fig. \ref{nv1}.

We consider the position of $S_{i,j}$ is located at $[x_{i,j},y_{i,j},z_{i,j}]$. 
Similar to the method used to obtain \eqref{sd01} in \eqref{xvs}, the considered intra-sector interference is well approximated in \eqref{sd1}.
\begin{figure*}[!t]
	\normalsize
	\begin{align}
		\label{sd1}
		\mathbb{I}_{j,\text{intra}} = P_{i,j} h_L({L_{i,j}})   
		G_n G_{\textrm{max}S_{i,1}} 
		\frac{1-\cos\left( N'_{i,1} k d_a 
			\tan^{-1}\left(\sqrt{\tan^2(\theta_{xU_{i,j}}-\theta_{xU_{i,1}})+\tan^2(\theta_{yU_{i,j}}-\theta_{yU_{i,1}})}\right)	
			\right)}
		{{N'_{i,1}}^2 \left(\tan^{-1}\left(\sqrt{\tan^2(\theta_{xU_{i,j}}-\theta_{xU_{i,1}})+\tan^2(\theta_{yU_{i,j}}-\theta_{yU_{i,1}})}\right)\right)^2}.
	\end{align}
	\hrulefill
\end{figure*}
%
As shown in Fig. \ref{xv2}, the parameters $\theta_{xU_{i,j}}$ and $\theta_{yU_{i,j}}$ are the directions of transmitted signal by $S_{i,j}$ in the $x-z$ and $y-z$ Cartesian coordinates, respectively. Let $w_{i,j}$ denotes the bandwidth dedicated to $S_{i,j}$ to $A_{r_{i,j}}$ uplink. Now, the total intra-sector interference is
\begin{align}
	\label{sd2}
	\mathbb{I}_\textrm{intra} = \sum_{j=2}^{N_{SU_i}} \mathbb{I}_{j,\text{intra}} B_{i,j},
\end{align}
where $0\leq B_{i,j} \leq 1$  and relates to the intended spectrum allocation technique.
Notice that we have $\sum_{j=1}^{N_{SU_i}} B_{i,j} = R_u$ and for $R_u=1$ the intra-sector interference tends to zero. Because in this case, each intra-sector uplink will be dedicated a separate frequency band.





\subsection{Inter-Sector Interference Analysis}
Due to the higher probability of LoS of UAVs, any $S_{i',j}$ to $U_{i'}$ uplink that uses frequency band $w_1$ can interfere with the considered $S_{i,1}$ to $U_i$ uplink. For example, as graphically shown in Fig. \ref{nv2}, for the uplink communication at the $i'$th sector between $S_{i',j}$ to $U_{i'}$, the side lobes of $S_{i',j}$ cause interference with the considered $S_{i,1}$ to $U_i$ uplink at the $i$th sector. 
We consider the positions of $S_{i,j}$ is located at $[x_{i,j},y_{i,j},z_{i,j}]$.
Similar to the method used to obtain \eqref{sd01} in \eqref{xvs}, the considered intra-sector interference is well approximated in \eqref{sd3}.
\begin{figure*}[!t]
	\normalsize
	\begin{align}
		\label{sd3}
		\mathbb{I}_{i,j,\text{intra}} &= P_{i',j} h_L({L_{i',j}})   
		\frac{1-\cos\left( N_{{i',j}} k d_a \tan^{-1}\left(\sqrt{\tan^2(\theta'_{xU_{i',j}}-\theta_{xU_{i,1}})+\tan^2(\theta'_{y'U_{i,j}}-\theta_{yU_{i,1}})}\right)\right)}
		{N_{{i',j}}^2 \left(\tan^{-1}\left(\sqrt{\tan^2(\theta'_{xU_{i',j}}-\theta_{xU_{i,1}})+\tan^2(\theta'_{y'U_{i,j}}-\theta_{yU_{i,1}})}\right)\right)^2} \nonumber \\
		&~~~\times \frac{1-\cos\left( N'_{{i,1}} k d_a \tan^{-1}\left(\sqrt{\tan^2(\theta_{xU_{i',j}}-\theta_{xU_{i,1}})+\tan^2(\theta_{y'U_{i,j}}-\theta_{yU_{i,1}})}\right)\right)}
		{   {N'_{i,1}}^2 \left(\tan^{-1}\left(\sqrt{\tan^2(\theta_{xU_{i',j}}-\theta_{xU_{i,1}})+\tan^2(\theta_{y'U_{i,j}}-\theta_{yU_{i,1}})}\right)\right)^2}, 
	\end{align}
	\hrulefill
\end{figure*}
In \eqref{sd3}, the parameters $\theta'_{xU_{i',j}}$ and $\theta'_{yU_{i',j}}$ are the directions of the transmitted signal by $S_{i',j}$ in the $x-z$ and $y-z$ Cartesian coordinates, respectively.
Considering the blockage effect, the total inter-sector interference is formulated as
\begin{align}
	\label{sd4}
	\mathbb{I}_\textrm{inter} = \sum_{\substack{i'=1 \\ i'\neq i}}^{N_D}
	\sum_{j=1}^{N_{SU_i}} \mathbb{I}_{i',j,\text{inter}} B_{i',j} P_\textrm{LoS}(\theta_{\textrm{elev}i',j}),
\end{align}
where $\theta_{\textrm{elev}i',j}$ is the elevation angle of $S_{i'j}$ compared to $U_i$.

\subsection{Analytical Derivations}
Next, we first develop a tractable closed-form expression for the the end-to-end SINR of the considered uplink. Then, for ease of performance analysis, closed-form expressions for outage probability and channel capacity are provided. 

{\bf Theorem 1.}
{\it The PDF of end-to-end SINR of the considered uplink can be well modeled as
\begin{align}
	\label{pr1}
	&f_{\Gamma_{i,1}} (\Gamma_{i,1}) =   
	\mathbb{M}\left(\frac{\mu_{xy}}{\sigma_\theta}, \frac{\theta_m}{M\sigma_\theta}\right)     
	\delta\left(\Gamma_{i,1} - \frac{\mathbb{D}_1 \left( {N'_{i,1} k d_a  }   \right)^2} { 4 } \right) 
	 \nonumber \\  
	&+\sum_{m=1}^M  \left( \mathbb{M}\left(\frac{\mu_{xy}}{\sigma_\theta}, \frac{m\theta_m}{M\sigma_\theta}\right)
	-\mathbb{M}\left(\frac{\mu_{xy}}{\sigma_\theta}, \frac{(m+1)\theta_m}{M\sigma_\theta}\right)    \right) \nonumber\\
	& \textcolor{white}{\frac{\frac{\frac{1}{1}}{1}}{1}}\!\!\!\!\!\!\!\!
	\times 
	\delta\left( \Gamma_{i,1}- \frac{ \mathbb{D}_1 M^2   \sin^2 \left( \frac{N'_{i,1} k d_a m \theta_m}   {2M} \right)} { m^2 } \right)    
\end{align}
where $\delta(\cdot)$ is the well-known Dirac delta function and $\mathbb{D}_1$ is a function of the main channel parameters such as $N_{SU_i}$, $N_D$, $P_{i,j}$, $L_{i,j}$, $R_u$, $\theta_{\textrm{elev}i',j}$, $N_{i,j}$, and $N'_{i,1}$ which is computed as
\begin{align}
	\label{pr2}
	&\mathbb{D}_1\left( N_{SU_i}, N_D, P_{i,j}, L_{i,j}, R_u, \theta_{\textrm{elev}i',j}, N_{i,j}, N'_{i,1}  \right) = \nonumber \\
	&\left[  \textcolor{white}{\frac{\frac{\frac{\frac{\frac{1}{1}}{1}}{1}}{1}}{1}}\!\!\!\!\!\!\! \sum_{j=2}^{N_{SU_i}}          
	\frac{ D_j  
		\sin^2\left( \frac{N'_{i,1} k d_a \theta_{U_{i,j}}     B_{i,j}  }  {2}	\right)}
	{ \left(\theta_{U_{i,j}}\right)^2} 
	+\frac{ {N'_{i,1}}^2 \sigma_n^2  } {2} +
	\right.  \nonumber \\
	&\sum_{\substack{i'=1 \\ i'\neq i}}^{N_D}
	\sum_{j=1}^{N_{SU_i}} 
	\frac{  D_{i',j}
		\sin^2\left( \frac{N_{{i',j}} k d_a \theta'_{U_{i',j}}  }{2}\right)
		\sin^2\left( \frac{N'_{{i,1}} k d_a \theta_{U_{i',j}}  }  {2}\right)}
	{    {N_{i',j}}^2   \left(\theta'_{U_{i',j}}\theta_{U_{i',j}}\right)^2  } 
	\left. \textcolor{white}{\frac{\frac{\frac{\frac{\frac{1}{1}}{1}}{1}}{1}}{1}}\!\!\!\!\!\!\!\!\!\!\!\right]^{-1} \nonumber \\
	&    
	\times P_{i,1} h_L({L_{i,1}}) G_n G_{\textrm{max}S_{i,1}},
\end{align}
In \eqref{pr1},  $\mu_{xy}=\sqrt{\mu_x^2 + \mu_y^2}$, $D_j=P_{i,j} h_L({L_{i,j}})   G_n G_{\textrm{max}S_{i,1}}$, and
$D_{i',j}= 2 P_{i',j} h_L({L_{i',j}})     B_{i',j} P_\textrm{LoS}(\theta_{\textrm{elev}i',j})$.
Also, $\mathbb{M}(a,b)$ is the Marcum {\it Q}-function and can be formulated as
\begin{align}
	\label{op2}
	\mathbb{M}(a,b) = \int_b^\infty x \exp\left(-\frac{x^2+a^2}{2} \right) I_0(ax).
\end{align}}

\begin{IEEEproof}
	Please refer to Appendix \ref{AppA}.
\end{IEEEproof}

The accuracy of the proposed SINR distribution based on Dirac Delta function depends on the parameters $\theta_m$ and $M$. In the next section, by comparing with Monte-Carlo simulation results, we will check the accuracy of \eqref{pr1}. For large values of 
$\theta_m$ and when $M$ increases to infinity, the closed-form expression provided in \eqref{pr1} tends to the SINR distribution obtained by Monte-Carlo simulation. Even though the accuracy of \eqref{pr1} increases by increasing $M$ and $\theta_m$ at the cost of more processing load. In the next section, we provide the minimum values for $M$ and $\theta_m$ for which \eqref{ro2} offers an accurate approximation of \eqref{ro}. 

The proposed SINR distribution based on Dirac Delta function is a function of the important channel parameters which are listed in Table I. 
Although the appearance of \eqref{pr1} seems a bit complex, the proposed channel PDF consists of a series of simple addition and multiplication operators along with the Marcum {\it Q}-function. Note that the Marcum {\it Q}-function is an standard function which can be readily computed.

In some practical cases, the angular offset of UAV's antenna fluctuations is negligible. Under this condition, \eqref{pr1} can be simplified as follow.

{\bf Lemma 1.} {\it The PDF of the end-to-end SINR of the considered uplink can be well modeled as
\begin{align}
	\label{pr3}
	&f_{\Gamma_{i,1}} (\Gamma_{i,1}) =  
	\left[ 
	\exp\left(-\frac{\theta_m^2}{2M^2\sigma_\theta}^2\right)     
	\delta\left(\Gamma_{i,1} - \frac{\mathbb{D}_1 \left( {N'_{i,1} k d_a  }   \right)^2} { 4 } \right) 
	\right. \nonumber \\  
	&+\sum_{m=1}^M  \left( \exp\left(-\frac{m^2\theta_m^2}{2M^2\sigma_\theta^2}\right)
	-\exp\left(- \frac{(m+1)^2\theta_m^2}{2M^2\sigma_\theta^2}\right)    \right) \nonumber\\
	&\left. \textcolor{white}{\frac{\frac{\frac{1}{1}}{1}}{1}}\!\!\!\!\!\!\!\!
	\times 
	\delta\left( \Gamma_{i,1}- \frac{ \mathbb{D}_1 M^2   \sin^2 \left( \frac{N'_{i,1} k d_a m \theta_m}   {2M} \right)} { m^2 } \right)   
	\right].
\end{align}}
\begin{IEEEproof}
Please refer to Appendix \ref{AppA}.
\end{IEEEproof}
As we see, \eqref{pr3} consists of only a series of simple addition and multiplication operators.

Outage probability and channel capacity are the most popular metrics for wireless network efficiency and for characterizing system performance.
Next, closed-form expressions for the outage probability and channel capacity of the considered wireless network are provided. 

{\bf Lemma 2.} {\it For a given end-to-end SINR threshold $\Gamma_\textrm{th}$, outage probability is obtained as
\begin{align}
	\label{rw2}
	\mathbb{P}_\textrm{out} &= 
	\sum_{m=1}^M  \mathbb{D}_{1,m} \left( \mathbb{M}\left(\frac{\mu_{xy}}{\sigma_\theta}, \frac{m\theta_m}{M\sigma_\theta}\right)   \right. \nonumber \\
	&~~~~~~~~~~~~~~~~~
	\left.-\mathbb{M}\left(\frac{\mu_{xy}}{\sigma_\theta}, \frac{(m+1)\theta_m}{M\sigma_\theta}\right)    \right),
\end{align}
where
\begin{align}
	\mathbb{D}_{1,m} =  \left\{
	\begin{array}{rl}
		0& ~~ {\rm for}~~ \mathbb{D}_1 m^{-2}M^2   \sin^2 \left( \frac{N'_{i,1} k d_a m \theta_m}   {2M} \right)\geq \Gamma_\textrm{th}, \\
		1& ~~ {\rm for}~~ \mathbb{D}_1 m^{-2}M^2   \sin^2 \left( \frac{N'_{i,1} k d_a m \theta_m}   {2M} \right)< \Gamma_\textrm{th}. \\
	\end{array} \right. 
\end{align}}
\begin{IEEEproof}
Outage probability is defined as the point at which the end-to-end SINR falls below the threshold which is mathematically defined as \begin{align}
\label{rw1}
\mathbb{P}_\textrm{out} = \textrm{Prob}\left\{ \Gamma_{i,1}<\Gamma_\textrm{th} \right\}.
\end{align}
where $\Gamma_\textrm{th}$ is the SINR threshold defined to guarantee the requested quality of service. 
Using \eqref{pr1} and \eqref{rw1}, and after some manipulations, outage probability is derived in \eqref{rw2}.
\end{IEEEproof}


{\bf Lemma 3.} {\it The Shannon (ergodic) channel capacity of the considered uplink as a function of important channel parameters is obtained as 
\begin{align}
	\label{rw5}
	& \bar{\mathbb{C}_i} = \frac{w_\textrm{ma}  R_u}{N_{{SU}_i}}
	\mathbb{M}\left(\frac{\mu_{xy}}{\sigma_\theta}, \frac{\theta_m}{M\sigma_\theta}\right)
	\left[ \log\left(1+ \frac{\mathbb{D}_1 \left( {N'_{i,1} k d_a  }   \right)^2} { 4 } \right) \right.\nonumber \\
	&+\frac{w_\textrm{ma}  R_u}{N_{{SU}_i}} \sum_{m=1}^M  \left( \mathbb{M}\left(\frac{\mu_{xy}}{\sigma_\theta}, \frac{m\theta_m}{M\sigma_\theta}\right)
	-\mathbb{M}\left(\frac{\mu_{xy}}{\sigma_\theta}, \frac{(m+1)\theta_m}{M\sigma_\theta}\right)    \right)  \nonumber \\
	&\times \left. \log\left(1+ \frac{ \mathbb{D}_1 M^2   \sin^2 \left( \frac{N'_{i,1} k d_a m \theta_m}   {2M} \right)} { m^2 } \right)
	\right].
\end{align}}
\begin{IEEEproof}
Ergodic capacity assumes that the fading transitions through all possible fading states, and thus might not be very useful in practice for source transmission with fixed delay constraints.

In our considered system model, under the practical assumption that the channel state information is not available at the transmitter, the source data is transmitted at a constant rate. Therefore the Shannon (ergodic) capacity is given by \cite{goldsmith2005wireless}
\begin{align}
    \label{rw4}
	\bar{\mathbb{C}_i} = \frac{w_\textrm{ma}}{N_{{SU}_i}}\times R_u \int_0^\infty \log\left(1+ \Gamma_{i,1}\right)
	f_{\Gamma_{i,1}} (\Gamma_{i,1})  \rm{d}\Gamma_{i,1}. 
\end{align}
\end{IEEEproof}
Using \eqref{pr1} and \eqref{rw4}, and after some manipulations, the closed-form expression for ergodic channel capacity of the considered uplink is derived in \eqref{rw5}.

\section{Numerical and Simulation Results}

For our simulations, we consider standard values for system parameters, as follows. The carrier frequency $f_c=60$ GHz,  outage threshold $\mathbb{P}_\textrm{out,th}=10^{-3}$, SINR threshold $\Gamma_\textrm{th}=10$ dB, constant parameters ($\alpha=9.61,\beta=0.16$), and all SBSs have the same transmission power equal to $P_{i,j}=30$ dBm, and same antenna pattern $N_{i,j}=18$.
Moreover, we consider a square geographical area with $A_s=5\times5~\text{km}^2$ consisting of 25 square sectors with $A_{s_i}=1\times1~\text{km}^2$ for $i\in\{1,2,...,25\}$. Each sector consists 12 SBSs that are uniformly distributed. In the center of each sector, we consider a hovering UAV acting as a NFP where $h_{U_i}$ is a uniform RV in the range of 100-150 m. We randomly select an SBS from the central sector and then we want to study the uplink channel between the selected SBS to the central sector NFP under the influence of UAV fluctuations as well as the interference effects of other SBSs.
Now, we compute the parameters $L_{i,j}$s, $L_{i',j}$s, $\theta_{xU_{i,j}}$, $\theta_{yU_{i,j}}$, $\theta_{xU_{i',j}}$, $\theta_{yU_{i',j}}$, $\theta'_{xU_{i',j}}$, $\theta'_{xU_{i',j}}$, and elevation angle $\theta_{\text{elev}i',j}$ based on the rotation and transformation matrix provided in \cite{3gppf}.
Then, based on a requested reuse frequency number $R_u$, we assign the same frequency bands to the SBSs with the highest angle difference that have the least intra-sector interference.
Note that the proposed channel framework as well as the provided analytical expressions are applicable for a wide range of comprehensive and realistic deployment of SBSs and the considered distribution of SBS is only an example of SBS distribution in order to highlight and demonstrate the relationship between the system parameters.

%
\begin{figure}
	\begin{center}
		\includegraphics[width=3.0 in]{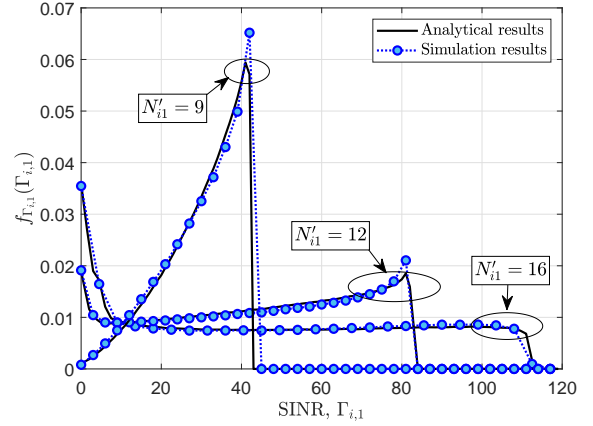}
		\caption{\textcolor{black}{Channel distribution of the considered UAV-based system for $\sigma_\theta=3^o$, $R_u=3$, and different values of $N'_{i,1}=9$, 12 and 16.}}
		\label{df1}
	\end{center}
\end{figure}
%

%
\begin{figure}
	\begin{center}
		\includegraphics[width=3.0 in]{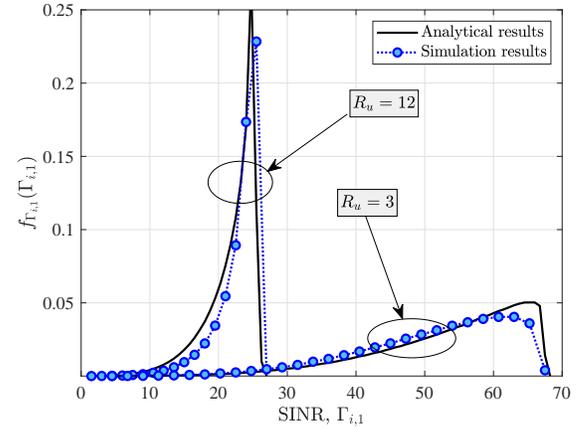}
		\caption{Channel distribution of the considered UAV-based system for $N'_{i,1}=10$, $\sigma_\theta=2^o$ and two different values of frequency reuse number $R_u=3$ and 12.}
		\label{df2}
	\end{center}
\end{figure}
%

First, by employing simulation results, the accuracy of the provided analytical channel distribution are corroborated under different channel conditions. 
As we observed in the previous section, the accuracy of the analytical expressions depends on two parameters $M$ and $\theta_m$. After a comprehensive search, it was found that $M=50$ and $\theta_m=\frac{6}{N'_{i,1}}$ are suitable values.
\textcolor{black}{In Fig. \ref{df1}, the channel distribution of the considered UAV-based system is plotted for different values of $N'_{i,1}=9$, 12 and 16. The results of Fig. \ref{df1} are obtained for $\sigma_\theta=3^o$ and $R_u=3$. As we observe, with an acceptable accuracy, the numerical results are close to the results obtained from simulations.
	The results clearly show that by changing the antenna pattern from $N'_{i,1}=9$ to 16, the SINR distribution changes significantly.
	In addition, from the results of Fig. \ref{df1}, a series of basic information about the performance of the considered system can be obtained. For instance, by increasing $N'_{i,1}$ from 9 to 16, the probability $\Gamma_{i,1}<10$\,dB increases. Therefore, by increasing $N'_{i,1}$, we expect the system performance to worsen in term of outage probability for the simulated conditions. On the other hand, as $N'_{i,1}$ increases, the probability of achieving larger values for $\Gamma_{i,1}$ increases, and thus, it seems that the channel capacity is increasing. However, we must note that the results about the system performance obtained from channel distribution are intuitive, and to give a more exact information, in the sequel, we comprehensively evaluate the system performance in terms of both channel capacity and outage probability under different channel conditions.} 
In addition to the antenna pattern, frequency reuse factor can also change the SINR distribution function. To show this, the channel distribution of the considered UAV-based system are depicted in Fig. \ref{df2} for two different values of frequency reuse number $R_u=3$ and 12. The results of Fig. \ref{df2} are obtained for $N'_{i,1}=10$ and $\sigma_\theta=2^o$. As we can see, same as for the antenna pattern, the reuse factor has a significant affect on the SINR distribution. In particular, in order to improve the the spectral efficiency of the considered system, we tend to increase $R_u$ as much as possible. However, by increasing $R_u$, the intra-sector interference increases, and thus, it decreases the end-to-end SINR at the receiver. In the following, the effect of $R_U$ on the system performance will be studied in more details.
Again, simulation results confirm the accuracy of the analytical derivations.

%
%

%
%

\begin{figure}
	\centering
	\subfloat[] {\includegraphics[width=3.1 in]{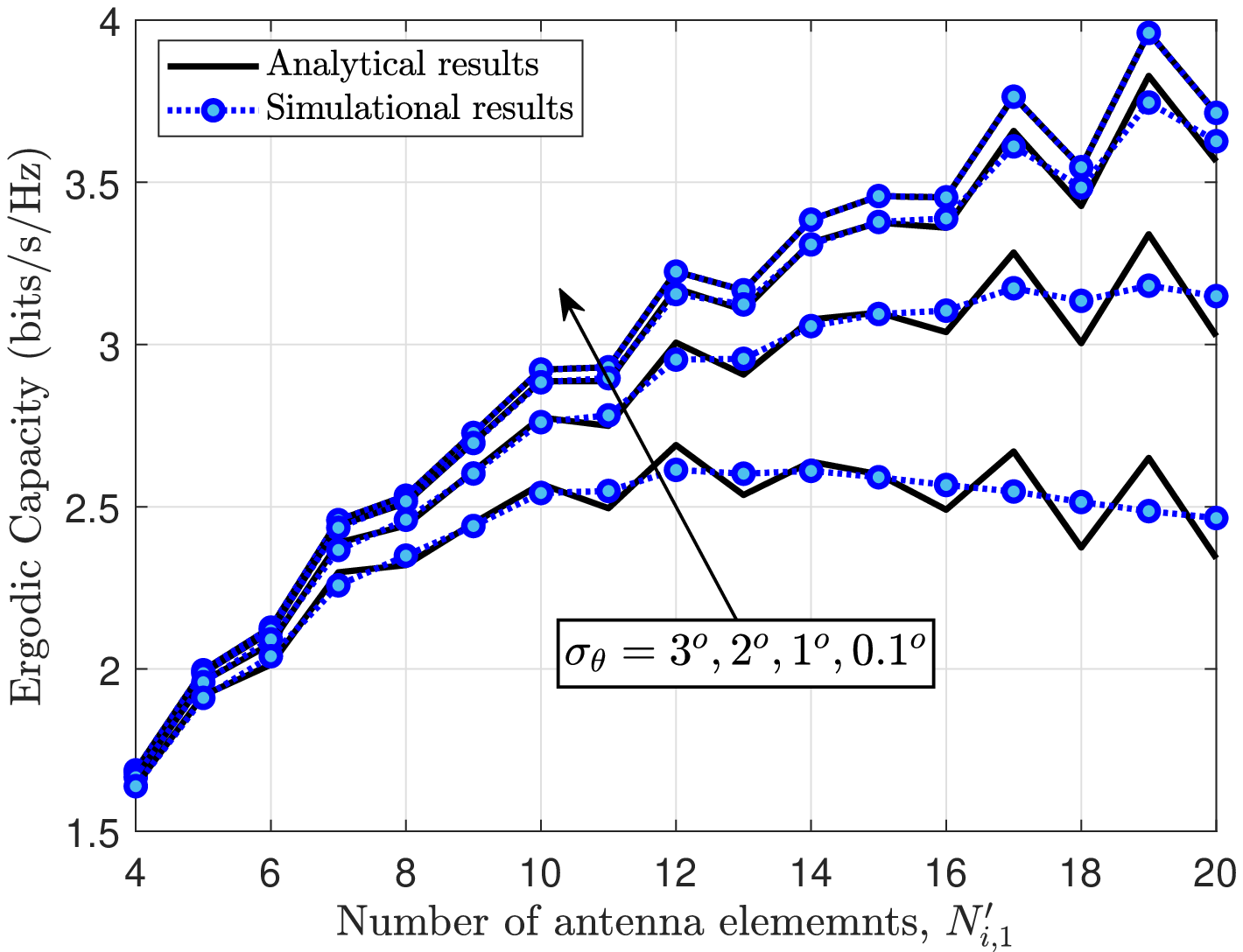}
		\label{df3}
	}
	\hfill
	\subfloat[] {\includegraphics[width=3.1 in]{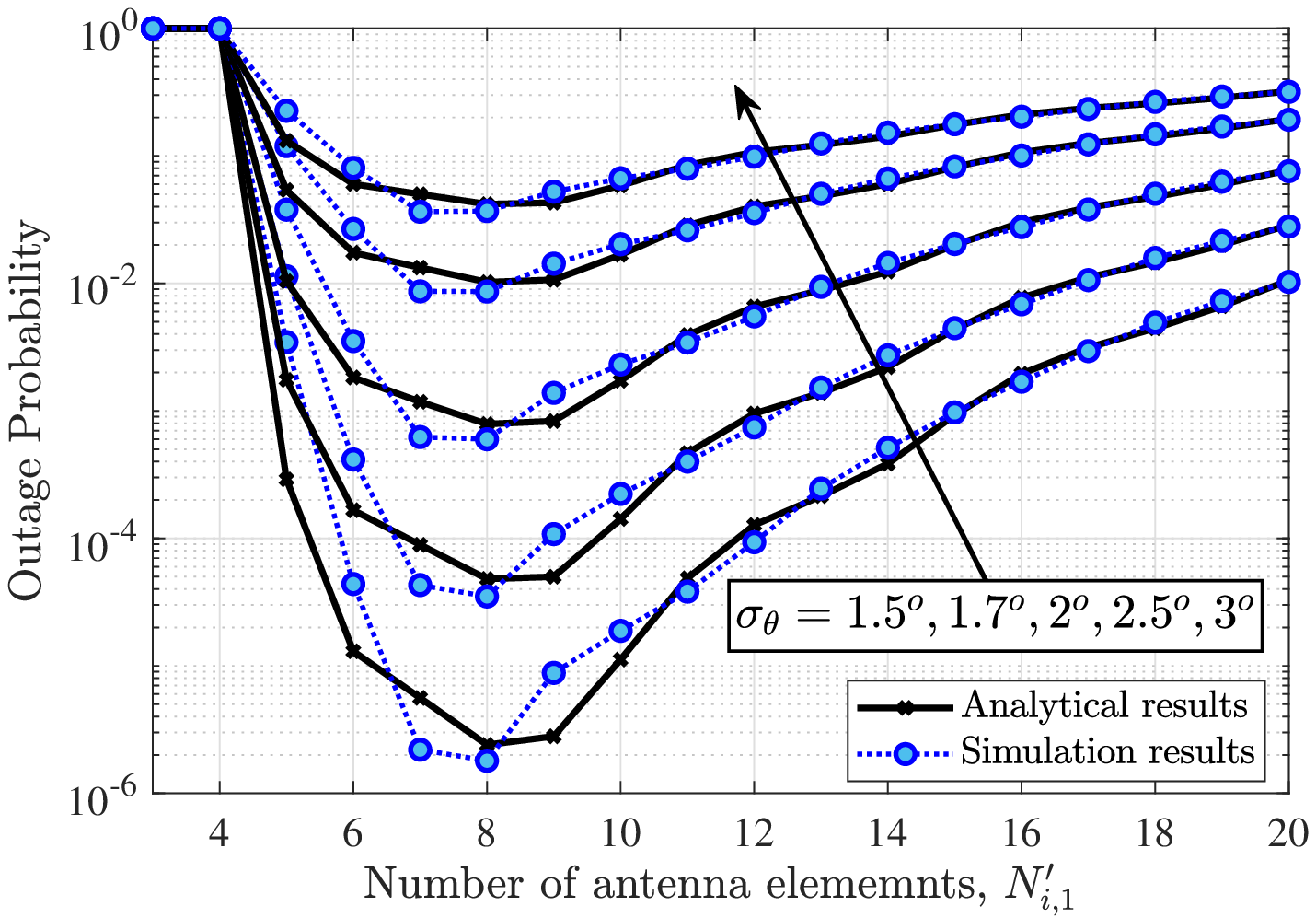}
		\label{df6}
	}
	\caption{(a) Ergodic capacity of the considered UAV-based system versus $N'_{i,1}$ for $R_u=6$ and different values of $\sigma_\theta=3^o,2^o,1^o,$ and $0.1^o$.
		(b) Outage probability of the considered UAV-based system versus $N'_{i,1}$ for $R_u=6$ and different values of $\sigma_\theta=3^o,2.5^o,2^o,1.7^o,$ and $01.5^o$.
	}
	\label{dff}
\end{figure}

Now, we want to investigate the effects of channel parameters such as $N'_{i,1}$, $\sigma_\theta$, and $R_u$ on the performance of the considered UAV-based uplink channel in terms of average capacity and outage probability. In Fig. \ref{df3}, ergodic capacity of the considered UAV-based system versus $N'_{i,1}$ are obtained for $R_u=6$ and a wide range of $\sigma_\theta=3^o,2^o,1^o,$ and $0.1^o$. Also, in Fig. \ref{df6}, we plot outage probability versus $N'_{i,1}$ for for $R_u=6$ and different values of $\sigma_\theta=3^o,2.5^o,2^o,1.7^o,$ and $01.5^o$.
As the antenna gain increases, both intra-sector and inter-sector interference decrease, and thus, the SINR at the receiver improves. As can be seen from Fig. \ref{df3}, it improves the system performance in terms of channel capacity.
However, as the antenna gain increases, the system becomes more sensitive to UAV instability. 
According to the results of Fig. \ref{df3}, by increasing the stability standard deviation of the UAV from $\sigma_\theta=1^o$ to $3^o$, it is observed that the channel capacity decreases from higher values of $N'_{i,1}$, while for lower values of $N'_{i,1}$, the instability has no significant effect on the channel capacity.
The results of Fig. \ref{df6} show that any changes in the intensity of UAV orientations have more effects on the system performance in term of outage probability compared to the channel capacity.

\begin{figure}
	\centering
	\subfloat[] {\includegraphics[width=3.0 in]{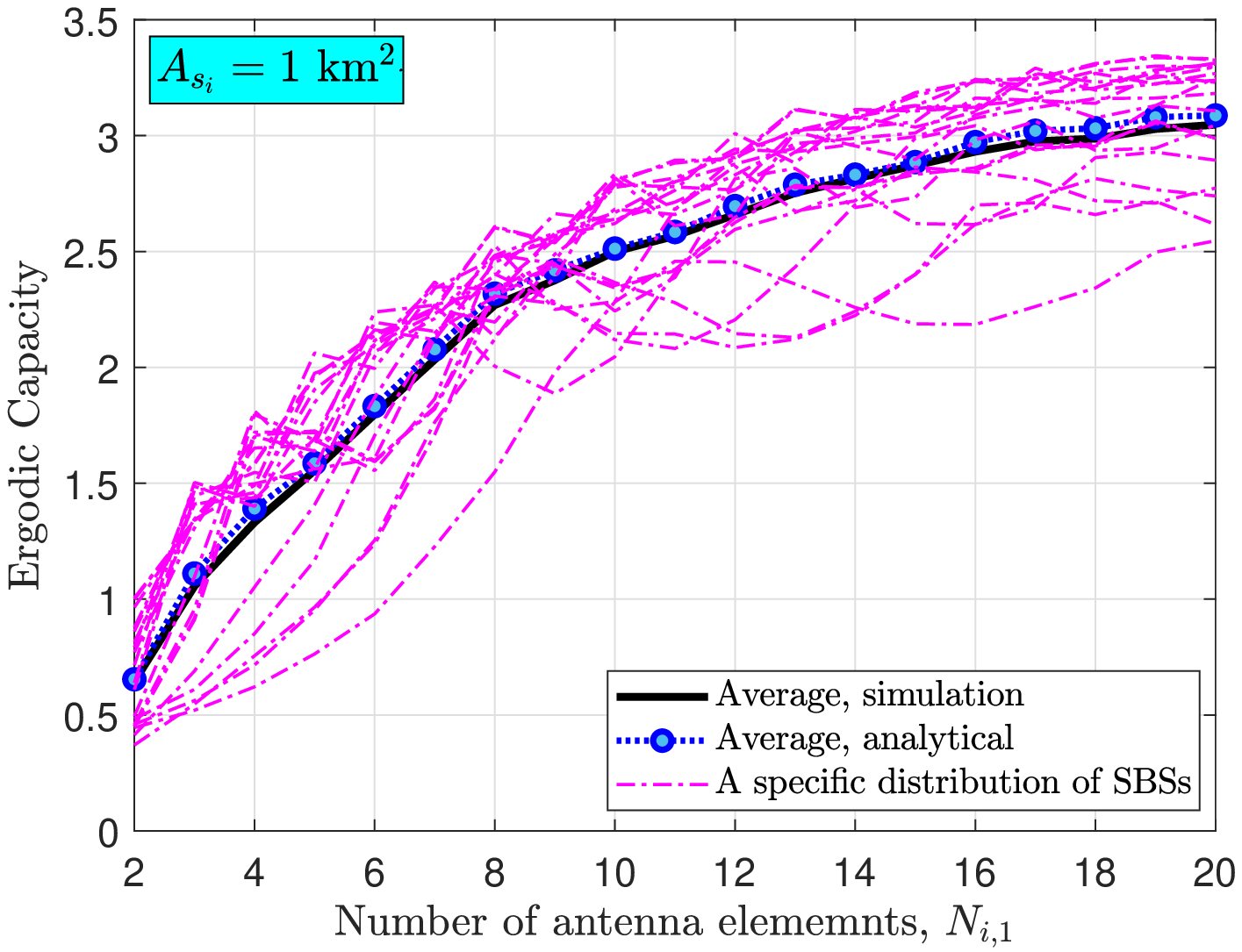}
		\label{mf1}
	}
	\hfill
	\subfloat[] {\includegraphics[width=3.0 in]{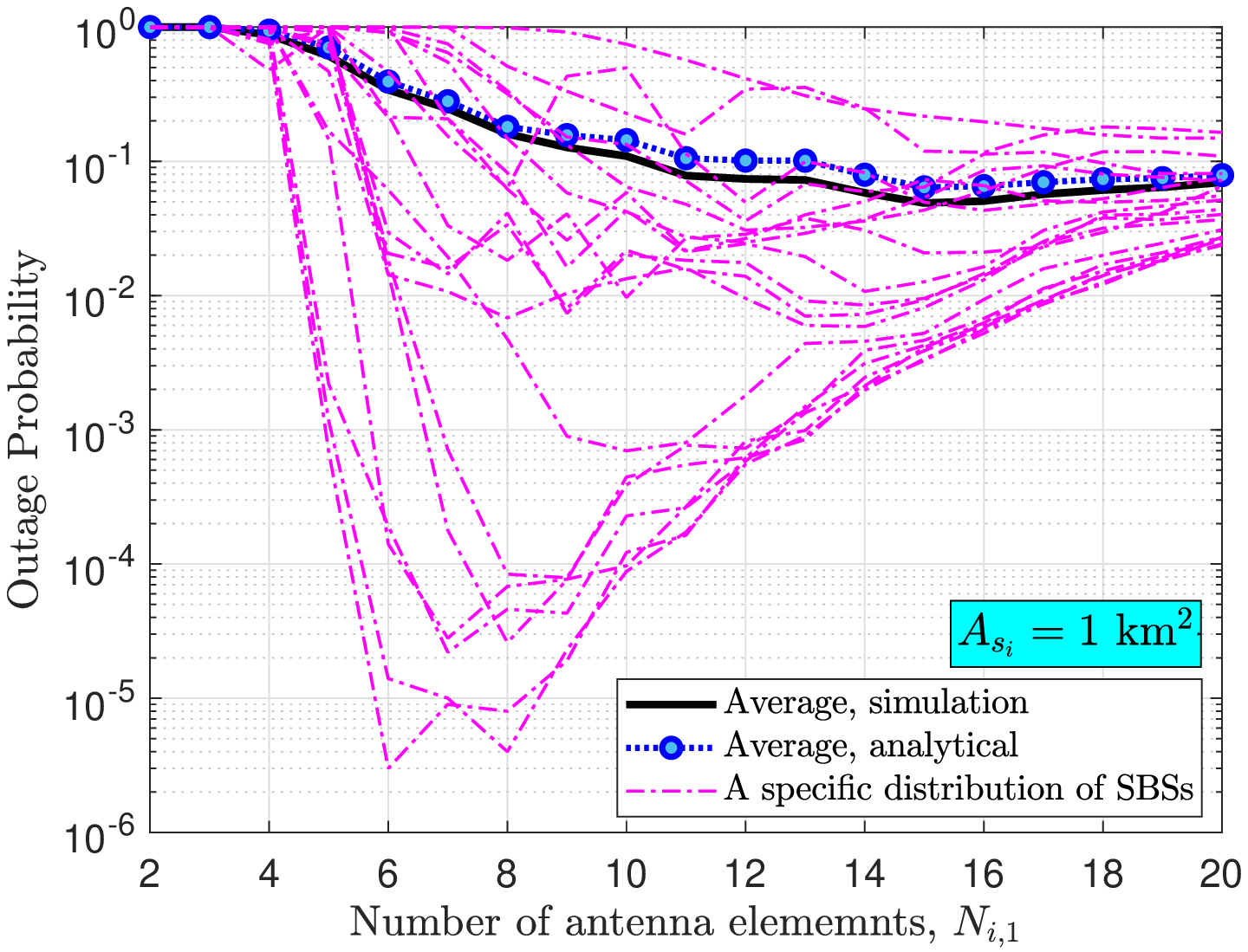}
		\label{mf2}
	}
	\caption{\textcolor{black}{
			Performance of the considered UAV-based system versus $N'_{i,1}$ for $R_u=6$ and 20 different distributions of SBSs in terms of (a) ergodic capacity, and (b) outage probability. In addition, the average values of ergodic capacity and outage probability are also provided.
	}}
	\label{mf3}
\end{figure}

\begin{figure}
	\centering
	\subfloat[] {\includegraphics[width=3.0 in]{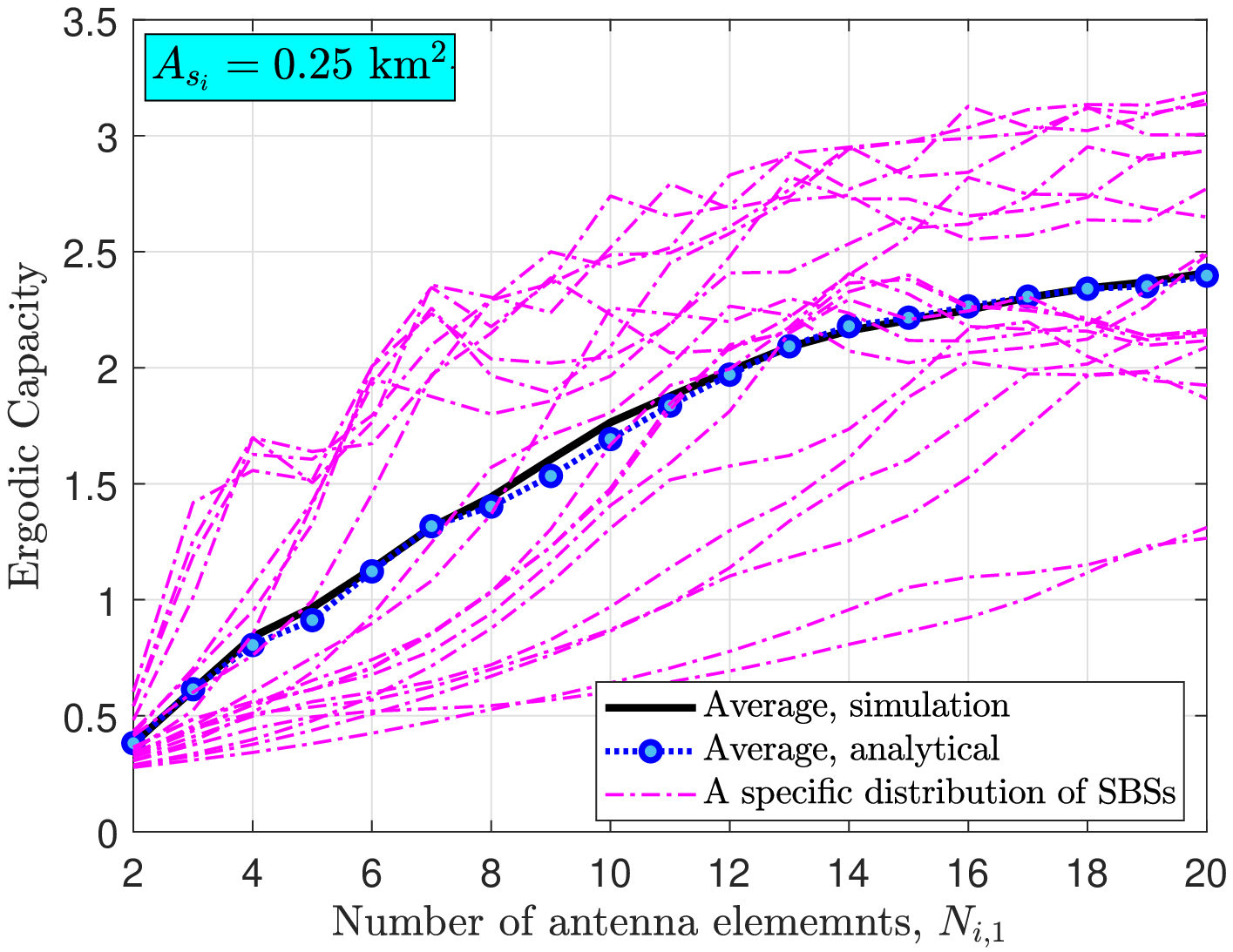}
		\label{nf1}
	}
	\hfill
	\subfloat[] {\includegraphics[width=3.0 in]{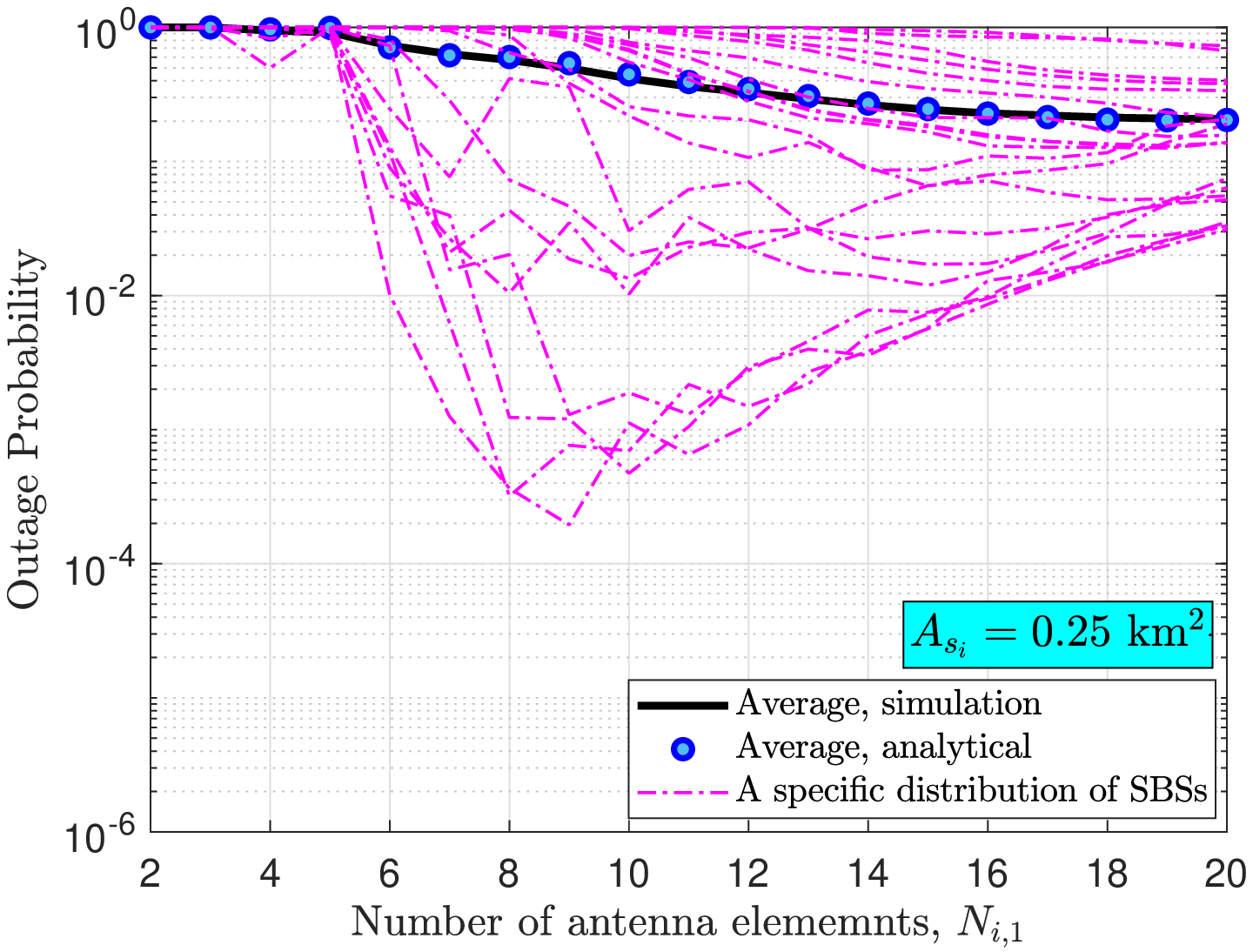}
		\label{nf2}
	}
	\caption{\textcolor{black}{
			Performance of the considered UAV-based system versus $N'_{i,1}$ for $A_{s_i}=0.25$ km$^2$, $R_u=6$ and 20 different distributions of SBSs in terms of (a) ergodic capacity, and (b) outage probability. In addition, the average values of ergodic capacity and outage probability are also provided.
	}}
	\label{nf3}
\end{figure}

\textcolor{black}{
	The distribution of SBSs is another important channel parameter affecting the system performance. To get a better insight, 
	we compare the system performance for 20 different random distributions of nodes.
	We randomly distribute the SBSs for 20 independent runs and compare the performance of these runs together in Fig. \ref{mf3}.
	Under these different distributions of SBSs, the channel capacity and outage probability are depicted versus $N'_{i,1}$ in Figs. \ref{mf1} and \ref{mf2}, respectively. 
	We expect that with each change in the distribution of SBSs, both intra-sector and inter-sector interference will change, resulting in a change in the system performance. The results of Fig. \ref{mf1} show that the channel capacity changes slightly by varying the distribution of SBSs. 
	It is more important to note that unlike channel capacity, interference has a significant effect on the considered system in term of outage probability.
	Therefore, as we see from the results of Fig. \ref{mf2}, that varying the distribution of SBSs has a significant effect on outage probability, especially at lower values of $N_{i,1}$.
	In addition, in Fig. \ref{mf3}, we provide the average outage probability and channel capacity of 20 different runs. Although the average channel capacity has a more expectable value, the average outage probability is closer to the larger values of 20 different runs. It can be justified because the lower values for the outage probability have very negligible effect on the average outage probability. To better understand this point, a numerical example is provided. For example, if the outage probability for the four independent cases are respectively equal to $10^{-1}$, $10^{-3}$, $10^{-4}$, and $10^{-5}$, the average outage probability is limited to the larger outage probability as $\frac{10^{-1}+10^{-3}+10^{-4}+10^{-5}}{4}=0.0253\simeq\frac{10^{-1}}{4} $, and smaller values of outage probability have no effect on the average outage probability. For example, if the values of $10^{-3}$, $10^{-4}$, and $10^{-5}$ decrease to $10^{-8}$, $10^{-9}$, and $10^{-10}$, respectively, the average outage probability decreases from 0.0253 to 0.02.
	The main reason for the increase in outage probability for some runs is that two SBSs with the same frequency are placed at short angular distances in which the performance is limited by the intra-sector interference. As the density of SBSs increases, with a higher probability that the SBSs will be located close together, and the performance decreases, especially, in term of outage probability. To illustrate this, in Fig. \ref{nf3}, we distribute the SBSs in a more dense network compared to Fig. \ref{mf3}. The values of system parameters used in Fig. \ref{nf3} are the same as those used in Fig. \ref{mf3} and only sector area decreases from $A_{s_i}=1$ to 0.25 km$^2$.
	It is clear that by changing the distribution of SBSs, the performance of the considered system changes significantly, therefore, comparing the performance of a random run of Fig. \ref{nf3} with a random run of Fig. \ref{mf3} does not seem logical. Thereby, in order to have a more fair comparison, the results of Fig. \ref{nf3} are compared with the results for 20 independent runs of the random distribution of SBSs.
	The results of these figures clearly show that by compressing the distribution space of SBSs, the performance of the system decreases, especially in term of outage probability. 
	Note that, it is necessary to pay attention to few points here. The results of these figures are obtained for $R_u=6$. Definitely, using smaller values for $R_u$ will make it possible to compress the network further.
	Moreover, the results of this work can be used in future research to find the optimal solutions for spectrum and power allocation, how to optimally connect SBSs to NFPs, and optimal NFPs positioning relative to any distribution of SBSs in order to increase performance of the considered system.}


\begin{figure}
	\centering
	\subfloat[] {\includegraphics[width=2.8 in]{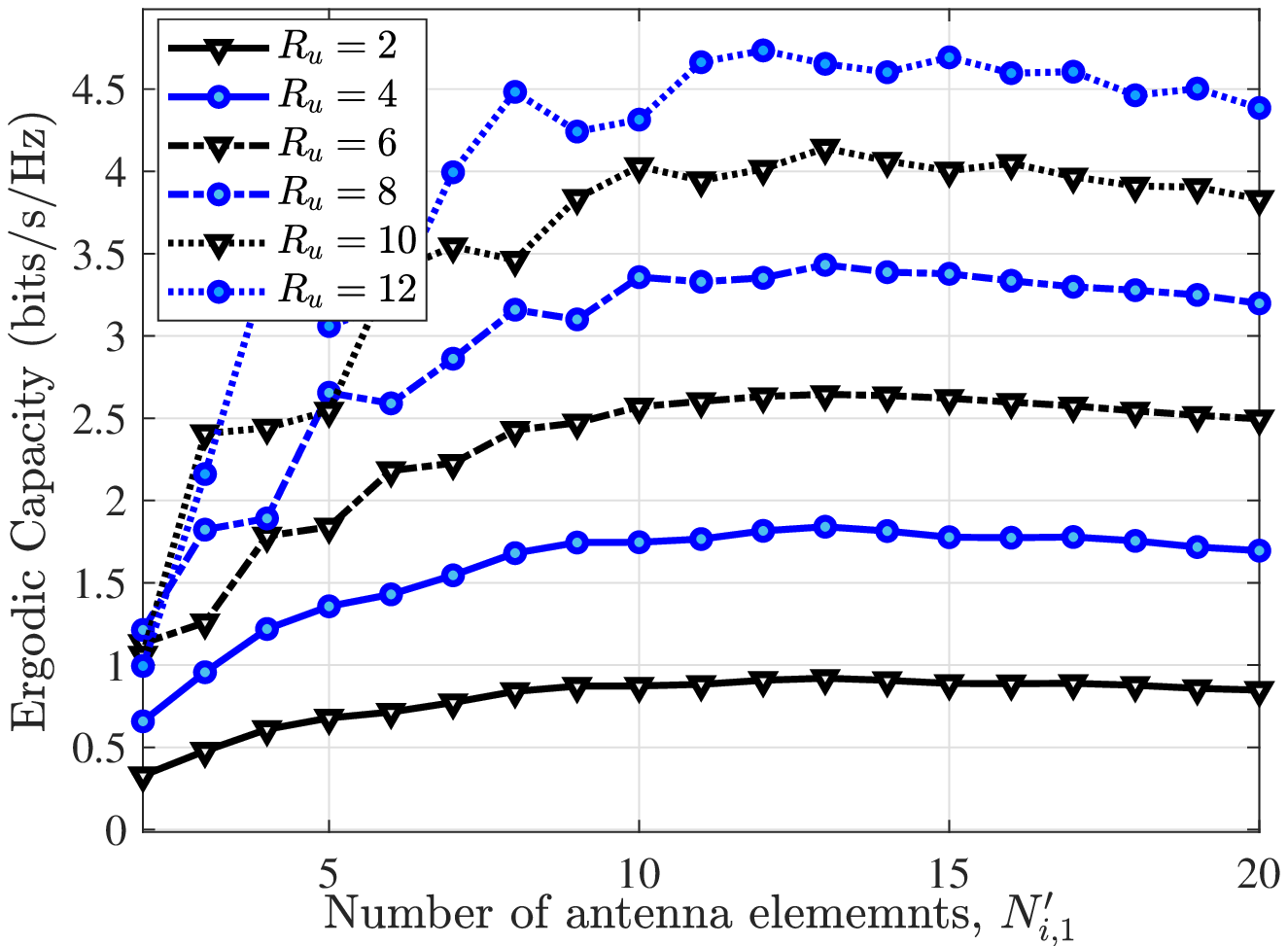}
		\label{mn1}
	}
	\hfill
	\subfloat[] {\includegraphics[width=2.8 in]{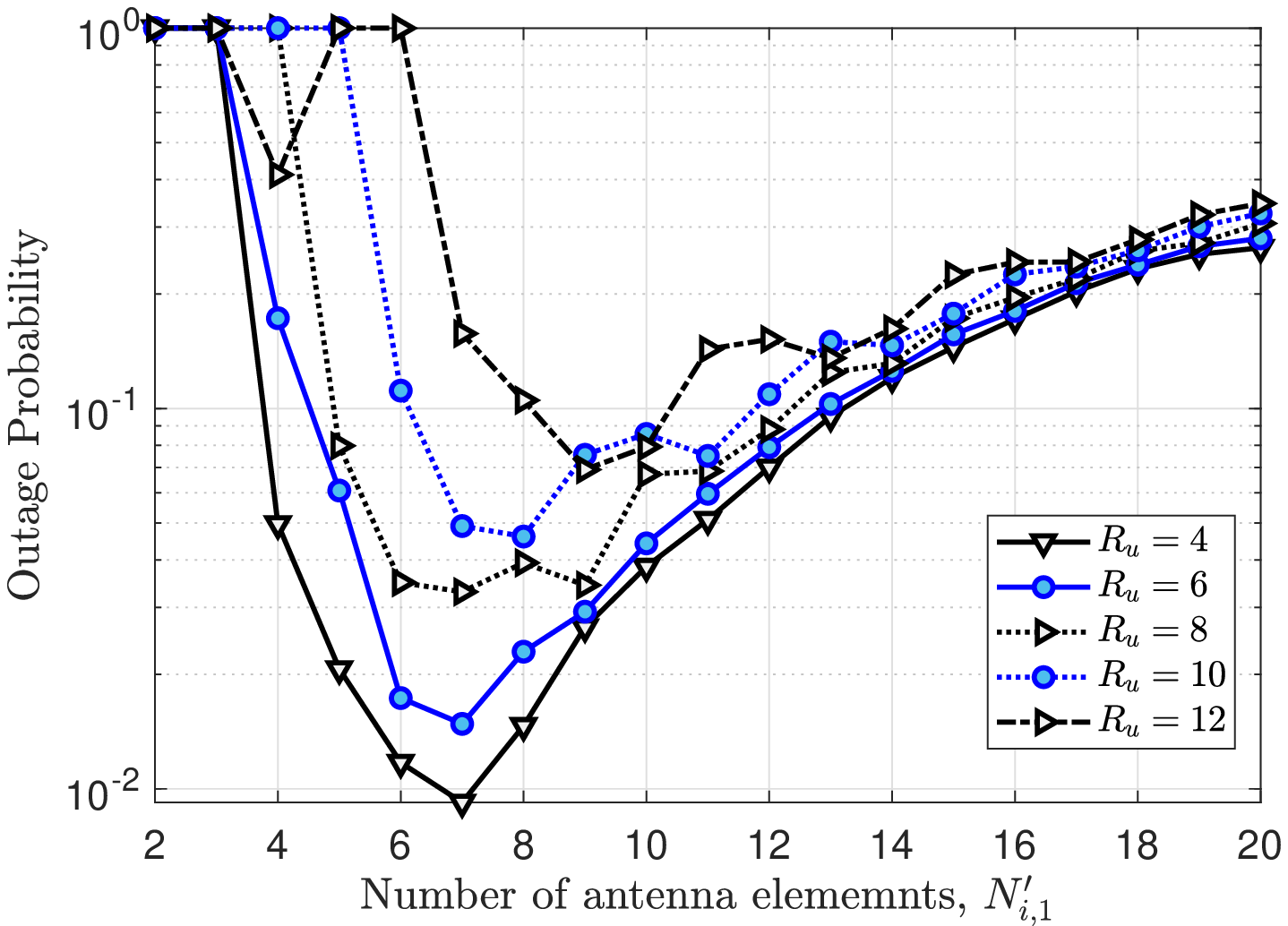}
		\label{mn2}
	}
	\caption{
		Performance of the considered UAV-based system versus $N'_{i,1}$ for different values of $R_u$ in terms of (a) ergodic capacity, and (b) outage probability.
	}
	\label{mn3}
\end{figure}

In Fig. \ref{mn3}, we investigate the effect of frequency reuse number $R_u$ on both outage probability and channel capacity. 
In order to increase the spectral efficiency of the network, we try to reuse the spectrum as much as possible, which allows us to assign a larger frequency band to each link, and as a result, the channel capacity improves. The results of Fig. \ref{mn1} clearly show that the channel capacity improves by increasing $R_u$. 
However, by increasing $R_u$, the distance between SBSs that use the same frequency band decreases and thus, the interference increases.
Therefore, as expected, the results of Fig. \ref{mn2} show that as $R_u$ increases, the performance of the system in the term of outage probability deteriorates.

Moreover, in Tables \ref{tab4} and \ref{tab5}, we want to find the optimal values for the channel parameters (such as $R_u$, $N'_{i,1}$).
As the simulation results show, the optimal values for the channel parameters are different in the terms of outage probability and channel capacity.
For optimal design of a communication network, we want to increase the transmission rate or channel capacity as much as possible while ensuring the quality of network services such as the maximum acceptable outage probability (or target outage probability). For instance, in Tables \ref{tab4} and \ref{tab5}, we consider that the target outage probability is $\mathbb{P}_\textrm{out,tr}=10^{-3}$.
The results of Tables \ref{tab4} and \ref{tab5} are obtained for $\sigma_\theta=1.7^o$ and $2.2^o$, respectively. Based on the results of Tables \ref{tab4}, and for the considered distribution of SBSs and $\sigma_\theta=1.7^o$, the maximum achievable channel capacity, that guarantees the target outage probability $\mathbb{P}_\textrm{out,tr}=10^{-3}$, is equal to $\bar{\mathbb{C}_i}=4.56$ bite/s/Hz, which is obtained for $N'_{1,i}=10$ and $R_u=10$. 
As the instability of the UAV increases, both intra-sector and inter-sector increase. Therefore, to reduce the interference effect and achieve maximum channel capacity, we must reduce the values of $N'_{i,1}$ and $R_u$.
For instance, based on the results of Table \ref{tab5}, when the instability of the UAV increases from $\sigma_\theta=1.7^o$ to $\sigma_\theta=2.2^o$, the maximum achievable channel capacity reduces to $\bar{\mathbb{C}_i}=2.75$ bite/s/Hz, which is obtained for lower values of $N'_{i,1}$ and $R_u$, i.e., $N'_{1,i}=8$ and $R_u=7$.
It is important to note that for a given geographical area, the stability of the UAV is affected by wind speed and with increasing wind speed, the instability of the UAV increases.
Since the coherence time of wind speed changes is usually in the order of a few minutes to a few hours, we expect that the variance of UAV's fluctuations also changes in the order of a few minutes to a few hours. 
Therefore, in order to achieve optimal performance, it is necessary to continuously calculate the optimal value of channel parameters such as $N'_{i,1}$ and $R_u$ in proportion to the instantaneous variance of the UAV's fluctuations.

%
\begin{table*}
\def\tablename{Table}
\centering
\caption{Comparison of the Optimal values for $N'_{1,i}$ and frequency reuse number to achieve maximum channel capacity as well as to guarantee an outage probability which is lower than a target threshold of $10^{-3}$ and for $\sigma_\theta=1.7^o$.
}
\begin{tabular}{|c||   
		c c|   c c|   c c|    c c|   c c|     c c|  c c| }
	\cline{1-15}  
	\cellcolor{Gray}  &
	\multicolumn{2}{|c|}{ \cellcolor{gray}$N'_{1,i}=4$ }   &  \multicolumn{2}{|c|}{ \cellcolor{gray} $N'_{1,i}=6$ }   &     
	\multicolumn{2}{|c|}{ \cellcolor{gray} $N'_{1,i}=8$ }  & 
	\multicolumn{2}{|c|}{ \cellcolor{green} $N'_{1,i}=10$ }  &  \multicolumn{2}{|c|}{ \cellcolor{gray} $N'_{1,i}=12$ }  &     
	\multicolumn{2}{|c|}{ \cellcolor{gray} $N'_{1,i}=14$ } & 
	\multicolumn{2}{|c|}{ \cellcolor{gray} $N'_{1,i}=14$ }  \\
	\cline{2-15}
	\cellcolor{Gray}$R_u$ 
	& 
	\cellcolor{lightgray}$\mathbb{P}_\textrm{out}$ & \cellcolor{lightgray} $\bar{\mathbb{C}_i}$ bps/Hz&      
	\cellcolor{lightgray} $\mathbb{P}_\textrm{out}$ & \cellcolor{lightgray} $\bar{\mathbb{C}_i}$ &
	\cellcolor{lightgray} $\mathbb{P}_\textrm{out}$ & \cellcolor{lightgray} $\bar{\mathbb{C}_i}$       &      
	\cellcolor{lightgray} $\mathbb{P}_\textrm{out}$ &\cellcolor{lightgray} $\bar{\mathbb{C}_i}$ &
	\cellcolor{lightgray} $\mathbb{P}_\textrm{out}$ &\cellcolor{lightgray} $\bar{\mathbb{C}_i}$       &      
	\cellcolor{lightgray} $\mathbb{P}_\textrm{out}$ &$\bar{\mathbb{C}_i}$ &
	\cellcolor{lightgray} $\mathbb{P}_\textrm{out}$ & \cellcolor{lightgray} $\bar{\mathbb{C}_i}$  \\
	\hline \hline 
	\cellcolor{Gray} 4& 
	\cellcolor{green}$<10^{-3}$&1.24    &               \cellcolor{green}$<10^{-3}$&  1.49  &              \cellcolor{green}$<10^{-3}$&  1.77  &          
	\cellcolor{green}$<10^{-3}$&  1.93  &              \cellcolor{green}$<10^{-3}$&  2.05  &               \cellcolor{orange}$>10^{-3}$& 2.19   &              
	\cellcolor{orange}$>10^{-3}$&   2.23      \\
	\hline
	\cellcolor{Gray} 6&  
	\cellcolor{orange}$>10^{-3}$& 1.83   &              \cellcolor{green}$<10^{-3}$& 2.28   &              \cellcolor{green}$<10^{-3}$& 2.59   &          
	\cellcolor{green} $<10^{-3}$& 2.82   &             \cellcolor{green}$<10^{-3}$& 3.00   &               \cellcolor{orange}$>10^{-3}$& 3.14   &              
	\cellcolor{orange}$>10^{-3}$&  4.23      \\
	\hline
	\cellcolor{Gray} 8&  
	\cellcolor{orange}$>10^{-3}$&  1.91  &               \cellcolor{green}$<10^{-3}$& 2.64   &              \cellcolor{green}$<10^{-3}$& 3.42  &          
	\cellcolor{green}$<10^{-3}$& 3.79   &        \cellcolor{green}$<10^{-3}$&  3.82  &               \cellcolor{orange}$>10^{-3}$& 3.98  &              
	\cellcolor{orange}$>10^{-3}$&   3.08      \\
	\hline
	\cellcolor{green} 10&  
	\cellcolor{orange}$>10^{-3}$&  2.47  &               \cellcolor{green}$<10^{-3}$& 3.64   &              \cellcolor{green}$<10^{-3}$& 3.95  &          
	\cellcolor{green}$<10^{-3}$& \cellcolor{green} 4.56   &        \cellcolor{orange}$<10^{-3}$&  4.65 &               \cellcolor{orange}$>10^{-3}$& 4.79  &              
	\cellcolor{orange}$>10^{-3}$&   5.33     \\
	\hline
	\cellcolor{Gray} 11&  
	\cellcolor{orange}$>10^{-3}$&  3.04  &              \cellcolor{orange} $>10^{-3}$&  3.37  &             \cellcolor{green} $<10^{-3}$&  4.08  &         
	\cellcolor{orange}$>10^{-3}$&  4.88  &              \cellcolor{orange} $>10^{-3}$&  4.79  &             \cellcolor{orange} $>10^{-3}$&  5.65  &              
	\cellcolor{orange}$>10^{-3}$&  5.36      \\
	\hline 
	\cellcolor{Gray} 12&  
	\cellcolor{orange}$>10^{-3}$&  3.60  &              \cellcolor{orange} $>10^{-3}$&  3.25  &             \cellcolor{orange} $>10^{-3}$&  5.00  &         
	\cellcolor{orange}$>10^{-3}$&  4.61  &              \cellcolor{orange} $>10^{-3}$&  5.67  &             \cellcolor{orange} $>10^{-3}$&  5.49  &              
	\cellcolor{orange}$>10^{-3}$&  6.12     \\
	\hline \hline
\end{tabular}
\label{tab4}%
\end{table*}%
%
%

%
\begin{table*}
	\def\tablename{Table}
	\centering
	\caption{Comparison of the Optimal values for $N'_{1,i}$ and frequency reuse number to achieve maximum channel capacity as well as to guarantee an outage probability which is lower than a target threshold of $10^{-3}$ and for $\sigma_\theta=2.2^o$.
	}
	\begin{tabular}{|c||   
			c c|   c c|   c c|    c c|   c c|     c c|  c c| }
		\cline{1-15}  
		\cellcolor{Gray}  &
		\multicolumn{2}{|c|}{ \cellcolor{gray} $N'_{1,i}=4$ }   &  \multicolumn{2}{|c|}{ \cellcolor{gray} $N'_{1,i}=6$ }   &     
		\multicolumn{2}{|c|}{ \cellcolor{green} $N'_{1,i}=8$ }  & 
		\multicolumn{2}{|c|}{ \cellcolor{gray} $N'_{1,i}=10$ }  &  \multicolumn{2}{|c|}{ \cellcolor{gray} $N'_{1,i}=12$ }  &     \multicolumn{2}{|c|}{ \cellcolor{gray} $N'_{1,i}=14$ } & 
		\multicolumn{2}{|c|}{ \cellcolor{gray} $N'_{1,i}=14$ }  \\
		\cline{2-15}
		\cellcolor{Gray}$R_u$ 
		& 
		\cellcolor{lightgray} $\mathbb{P}_\textrm{out}$ &\cellcolor{lightgray} $\bar{\mathbb{C}_i}$ bps/Hz&      
		\cellcolor{lightgray} $\mathbb{P}_\textrm{out}$ &\cellcolor{lightgray} $\bar{\mathbb{C}_i}$ &
		\cellcolor{lightgray} $\mathbb{P}_\textrm{out}$ &\cellcolor{lightgray} $\bar{\mathbb{C}_i}$       &      
		\cellcolor{lightgray} $\mathbb{P}_\textrm{out}$ &\cellcolor{lightgray} $\bar{\mathbb{C}_i}$ &
		\cellcolor{lightgray} $\mathbb{P}_\textrm{out}$ & \cellcolor{lightgray} $\bar{\mathbb{C}_i}$       &      
		\cellcolor{lightgray} $\mathbb{P}_\textrm{out}$ & \cellcolor{lightgray} $\bar{\mathbb{C}_i}$ &
		\cellcolor{lightgray} $\mathbb{P}_\textrm{out}$ &\cellcolor{lightgray} $\bar{\mathbb{C}_i}$  \\
		\hline \hline 
		\cellcolor{Gray} 4& 
		\cellcolor{orange}$>10^{-3}$&1.23    &               \cellcolor{green}$<10^{-3}$&  1.46  &              \cellcolor{green}$<10^{-3}$&  1.74  &          
		\cellcolor{orange}$>10^{-3}$&  1.87  &              \cellcolor{orange}$>10^{-3}$&  1.98  &               \cellcolor{orange}$>10^{-3}$& 2.13   &              
		\cellcolor{orange}$>10^{-3}$&   2.21      \\
		\hline
		\cellcolor{Gray} 5&  
		\cellcolor{orange}$>10^{-3}$& 1.37   &              \cellcolor{green}$<10^{-3}$& 1.85   &              \cellcolor{green}$<10^{-3}$& 2.21   &          
		\cellcolor{green} $<10^{-3}$& 2.32   &             \cellcolor{orange}$>10^{-3}$& 2.44   &               \cellcolor{orange}$>10^{-3}$& 2.51   &              
		\cellcolor{orange}$>10^{-3}$&  2.58      \\
		\hline
		\cellcolor{Gray} 6&  
		\cellcolor{orange}$>10^{-3}$&  1.81  &               \cellcolor{green}$<10^{-3}$& 2.24   &              \cellcolor{green}$<10^{-3}$&  2.54  &          
		\cellcolor{orange}$>10^{-3}$& 2.77   &        \cellcolor{orange}$>10^{-3}$&  2.89  &               \cellcolor{orange}$>10^{-3}$& 2.99   &              
		\cellcolor{orange}$>10^{-3}$&   3.08      \\
		\hline
		\cellcolor{green} 7&  
		\cellcolor{orange}$>10^{-3}$&  1.96  &               \cellcolor{orange}$>10^{-3}$& 2.41   &              \cellcolor{green}$<10^{-3}$& \cellcolor{green}2.75  &          
		\cellcolor{orange}$>10^{-3}$& 3.04   &               \cellcolor{orange}$>10^{-3}$&  3.28  &              \cellcolor{orange}$>10^{-3}$&  3.44  &              
		\cellcolor{orange}$>10^{-3}$&  3.61       \\
		\hline
		\cellcolor{Gray} 8&  
		\cellcolor{orange}$>10^{-3}$&  1.91  &              \cellcolor{orange} $>10^{-3}$&  2.63  &             \cellcolor{orange} $>10^{-3}$&  3.34  &         
		\cellcolor{orange}$>10^{-3}$&  3.65  &              \cellcolor{orange} $>10^{-3}$&  3.69  &             \cellcolor{orange} $>10^{-3}$&  3.82  &              
		\cellcolor{orange}$>10^{-3}$&  3.94      \\
		\hline \hline
	\end{tabular}
\label{tab5}%
\end{table*}%
%
%

\section{Conclusion}
In this work, first, we characterized an accurate mmWave SINR distribution for SBS to aerial NFP by taking into
consideration real parameters such as UAV's vibrations, distribution of SBSs, position of UAVs in the sky, real 3D antenna pattern model provided by 3GPP along with interference caused by antenna side lobes and frequency reuse. 
Then, for the characterized channel, we derived closed-form expressions for channel distribution, outage probability and channel capacity.
As we observed, the accuracy of the derived analytical expressions was verified by Monte Carlo simulations.
Next, we investigated the effects of channel parameters such as antenna pattern gain, strength of UAV's vibrations, UAVs' positions in the sky, distribution of SBSs, and frequency reuse on the performance of the considered UAV-based uplink channel in terms of average capacity and outage probability.
%
%
As we observed, by increasing antenna gain, both intra-sector and inter-sector interference decrease, and thus, the SINR at the receiver improves which enhances the system performance in terms of channel capacity.
However, as the antenna gain increases, the system becomes more sensitive to UAV instability and significantly affects the performance in term of outage probability. 
Moreover, for different distributions of SBSs, both intra-sector and inter-sector interference change, resulting in a change in the system performance, especially in term of outage probability.

%
%
%

\appendices

\section{}
\label{AppA}
From \eqref{po1}, \eqref{sd01}, \eqref{sd1}, and \eqref{sd3}, finding a closed-form analytical expression for the PDF of $\Gamma_{i,1}$ is very difficult, if not impossible. As we show in section of IV, \eqref{po1} can be well approximated with \eqref{ro}. 
%
%
\begin{figure*}[!t]
	\normalsize
	\textcolor{black}{
	\begin{align}
		\label{ro} 
		\Gamma_{i,1} \simeq 	
		\frac{
			\frac{    
				P_{i,1} h_L({L_{i,1}}) G_n G_{\textrm{max}S_{i,1}} 
				\sin^2 \left( \frac{N'_{i,1} k d_a \theta_{U_{i,1}} }   {2} \right)
			} {(\theta_{U_{i,1}})^2 ~{N'_{i,1}}^2}     
		}
		{      \left[ \sum_{j=2}^{N_{SU_i}}          
			\frac{ 2 D_j  
				\sin^2\left( \frac{N'_{i,1} k d_a \theta_{U_{i,j}}     B_{i,j}  }  {2}	\right)}
			{ {N'_{i,1}}^2 \left(\theta_{U_{i,j}}\right)^2} 
			+   \sum_{\substack{i'=1 \\ i'\neq i}}^{N_D}
			\sum_{j=1}^{N_{SU_i}} 
			\frac{  D_{i',j}
				\sin^2\left( \frac{N_{{i',j}} k d_a \theta'_{U_{i',j}}  }{2}\right)
				\sin^2\left( \frac{N'_{{i,1}} k d_a \theta_{U_{i',j}}  }  {2}\right)}
			{  {N'_{i,1}}^2    {N_{i',j}}^2   \left(\theta'_{U_{i',j}}\theta_{U_{i',j}}\right)^2  }
			+\sigma_n^2
			\right] }
	\end{align}}
	\hrulefill
\end{figure*}
In \eqref{ro}, $D_j=P_{i,j} h_L({L_{i,j}})   G_n G_{\textrm{max}S_{i,1}}$, and
$D_{i',j}= 2 P_{i',j} h_L({L_{i',j}})     B_{i',j} P_\textrm{LoS}(\theta_{\textrm{elev}i',j})$.
As we will show, by increasing $\theta_{U_{i,1}}$, the value of $\Gamma_{i,1}$ tends to zero. Therefore, for simplicity, we truncate $\Gamma_{i,1}$ as
\begin{align}
	\label{ro3}
	\Gamma_{i,1}(\theta_m) = \left\{
	\begin{array}{rl}
		\Gamma_{i,1}& ~~~ {\rm for}~~~ \theta_{U_{i,1}}\leq \theta_m \\
		0~~~& ~~~ {\rm for}~~~ \theta_{U_{i,1}}> \theta_m \\
	\end{array} \right. .
\end{align}
Now, by sectorizing \eqref{ro3} similar to the sectorized method used in \cite{dabiri2020analytical}, we obtain \eqref{ro2}.   
\begin{figure*}[!t]
	\normalsize
	\begin{align}
		\label{ro2}
		&\Gamma_{i,1}(\theta_m,M) \simeq 
		\left[ \sum_{j=2}^{N_{SU_i}}          
		\frac{ D_j  
			\sin^2\left( \frac{N'_{i,1} k d_a \theta_{U_{i,j}}     B_{i,j}  }  {2}	\right)}
		{ \left(\theta_{U_{i,j}}\right)^2} 
		+   \sum_{\substack{i'=1 \\ i'\neq i}}^{N_D}
		\sum_{j=1}^{N_{SU_i}} 
		\frac{  D_{i',j}
			\sin^2\left( \frac{N_{{i',j}} k d_a \theta'_{U_{i',j}}  }{2}\right)
			\sin^2\left( \frac{N'_{{i,1}} k d_a \theta_{U_{i',j}}  }  {2}\right)}
		{    {N_{i',j}}^2   \left(\theta'_{U_{i',j}}\theta_{U_{i',j}}\right)^2  }
		+\frac{ {N'_{i,1}}^2 \sigma_n^2  } {2}
		\right]^{-1}	
		\nonumber \\
		&\times
		%
		\left(  \frac{    
			\left( {N'_{i,1} k d_a  }   \right)^2
			\left[\Pi\left(\theta_{U_{i,1}}\right) - \Pi\left(\theta_{U_{i,1}}-\frac{\theta_m}{M}\right)   \right]
		} { 4 }
		+\sum_{m=1}^M \frac{    
			M^2   \sin^2 \left( \frac{N'_{i,1} k d_a m \theta_m }   {2M} \right)
			\left[\Pi\left(\theta_{U_{i,1}}-\frac{m\theta_m}{M}\right) - \Pi\left(\theta_{U_{i,1}}-\frac{(m+1)\theta_m}{M}\right)   \right]
		} { m^2 }  \right)  \nonumber \\
		&\times P_{i,1} h_L({L_{i,1}}) G_n G_{\textrm{max}S_{i,1}},     ~~~~~~~~~~~~~~~~~~~~~~~~~~~~~~~~~~~~~~~~~~~~~~~~~~
		\textrm{for}~~~~~~~0<\theta_{U_{i,1}}<\theta_m.
	\end{align}
	\hrulefill
\end{figure*}
In \eqref{ro2}, the parameter $M$ denote the number of sectors and 
$
\Pi(x)= \left\{
\begin{array}{rl}
	1& ~~~ {\rm for}~~~ x\geq 0 \\
	0& ~~~ {\rm for}~~~ x<1 \\
\end{array} \right. 
$
Note that, for large value of 
$\theta_m$ and when $M$ grows to infinity, \eqref{ro2} tends to \eqref{ro}. Even though the accuracy of the proposed sectorized model increases by increasing $M$ and $\theta_m$ at the cost of more processing load. We provide the minimum values for $M$ and $\theta_m$ for which \eqref{ro2} offers an accurate approximation of \eqref{ro}.

For higher UAV's orientation stability, we can approximate $\theta_{U_{i,1}}\simeq\sqrt{\theta_{xU_{i,j}}^2+\theta_{yU_{i,j}}^2}$.
Since we have $\theta_{xU_{i,j}}\sim \mathcal{N}(\mu_x,\sigma^2_{\theta})$, and $\theta_{yU_{i,1}}\sim \mathcal{N}(\mu_y,\sigma^2_\theta)$, then, the distribution of RV $\theta_{U_{i,1}}$ becomes Rician as
\begin{align}
	\label{fg1}
	f_{\theta_{U_{i,1}}}(\theta_{U_{i,1}}) = 
	\frac{\theta_{U_{i,1}}}{\sigma_\theta^2}
	\exp\left(-\frac{\theta_{U_{i,1}}^2+\mu_{xy}^2}{2\sigma_\theta^2} \right) 
	 I_0 \left( \frac{\theta_{U_{i,1}}\mu_{xy}}{\sigma_\theta^2}\right),
\end{align}
where $\mu_{xy} = \sqrt{\mu_x^2+\mu_y^2}$, and $I_0(.)$ is the modified Bessel function of the first kind with order zero.
From \eqref{ro2}, \eqref{fg1} and using \cite{math_wolfram}, after some mathematical manipulations, the distribution of $\Gamma_{i,1}$ is derived in \eqref{pr1}.

In some practical cases, the angular offset of UAV's antenna fluctuations is negligible. Under this condition, \eqref{fg1} is simplified as
\begin{align}
	\label{fg2}
	f_{\theta_{U_{i,1}}}(\theta_{U_{i,1}}) = 
	\frac{\theta_{U_{i,1}}}{\sigma_\theta^2}
	\exp\left(-\frac{\theta_{U_{i,1}}^2}{2\sigma_\theta^2} \right).
\end{align}
Following the method used to obtain \eqref{fg2}, and after some mathematical manipulations, the distribution of $\Gamma_{i,1}$ when $\mu_x\simeq\mu_y\simeq0$ is derived as in \eqref{pr3}.




\end{document}